\documentclass[aip, pof, floatfix, preprint]{revtex4-1}
\usepackage{fancyhdr}
\usepackage{siunitx}
\usepackage{graphicx}
\usepackage{subfigure} 
\usepackage{float} 
\usepackage{tabularx}
\usepackage{color}
\usepackage{geometry}
\usepackage{mciteplus}
\usepackage{footnote}
\usepackage{multirow}
\usepackage{textcomp}
\usepackage{color,soul}
\usepackage{lineno}
\usepackage{multirow}
\usepackage{amssymb}
\usepackage{amsmath}
\usepackage{mathtools}
\usepackage{commath}
\usepackage{natbib}
\usepackage[intoc]{nomencl}
\usepackage{array}
\usepackage{nomencl}
\makenomenclature
\usepackage{makecell}
\usepackage{graphicx}
\usepackage{caption}
\usepackage{array}
\usepackage{subcaption}
\usepackage{booktabs}
\usepackage{blindtext}
\usepackage{soul}
\usepackage{isomath}

\usepackage[version=3]{mhchem}

\makeatletter
\def\@email#1#2{%
	\endgroup
	\patchcmd{\titleblock@produce}
	{\frontmatter@RRAPformat}
	{\frontmatter@RRAPformat{\produce@RRAP{*#1\href{mailto:#2}{#2}}}\frontmatter@RRAPformat}
	{}{}
}%

\newcolumntype{P}[1]{>{\centering\arraybackslash}p{#1}}
\makenomenclature
 
\usepackage{etoolbox}
\renewcommand\nomgroup[1]{%
	\item[\bfseries
	\ifstrequal{#1}{A}{}{%
		\ifstrequal{#1}{G}{Greek Letters}{%
			\ifstrequal{#1}{S}{Subscripts}{}}}%
	]}
\makeatother

\begin{document}

	\preprint{}
	
	\title{Numerical investigation on solids settling in a non-newtonian slurry inside a horizontal flume}

	\author{Shubham Sharma}

	\affiliation{Department of Chemical and Materials Engineering, University of Alberta, Alberta T6G 1H9, Canada}

 \author{Somasekhara Goud Sontti}
 \affiliation{Multiphase Flow and Microfluidics(MFM)Laboratory,~Department of Chemical Engineering, Indian Institute of Technology Dharwad, Dharwad, Karnataka 580011, India}
  \author{Wenming Zhang }   
	\affiliation{Department of Civil and Environmental Engineering, University of Alberta, Edmonton, AB T6G 1H9, Canada}
    \author{Petr Nikrityuk}
	\affiliation{Department of Chemical and Materials Engineering, University of Alberta, Alberta T6G 1H9, Canada}
	\author{Xuehua Zhang}
	\affiliation{Department of Chemical and Materials Engineering, University of Alberta, Alberta T6G 1H9, Canada}
    \email{xuehua.zhang@ualberta.ca}
    \email{somasekhar.sonti@iitdh.ac.in}
	\date{\today}
	
	\setlength{\abovedisplayskip}{6pt}
	\setlength{\belowdisplayskip}{6pt}

\newpage
\begin{abstract}
Slurry transportation is always crucial for many industrial processes. This study numerically investigates the settling behavior of multisize solid particles in a non-newtonian slurry inside a semicircular open channel (flume). The non-newtonian slurry is modelled using a three-dimensional (3D) unsteady Eulerian-Eulerian (E-E) model coupled with the Hershel-Bulkley (HB) rheological model. A detailed sensitivity analysis of drag models is performed to establish the solid-fluid interaction in the slurry flow. The numerical model is validated with the experimental data from the literature and shows a fair agreement. The validated model is used to simulate the settling behavior of the slurry in the flume. The mean particle diameter of the solid particles in the slurry is in the range of 75-296 $\mu$m with a median diameter of $ 188 $ $\mu$m. The effect of particle size distributions (PSDs), flume inclination, bubble size and bubble volume fraction on the particle settling inside the flume is investigated in the parametric study. The analysis of our results revealed that the settling of solids is significantly affected by PSDs in the open channel system. In particular, the increase in flume inclination progresses the settling and dissipation of fine and coarse particles, respectively. Additional simulations showed that the inception of bubbles influences the settling velocity of solids, which changes the settling behavior of multisize solids inside the flume. The presented study can be used as a valuable guideline for the optimisation of intermediate exclusion of water from thickened slurry in order to ensure the stability of tailing storage facility.
\end{abstract}
\maketitle
\section{Introduction} 

Water-based bitumen extraction from oil sand always requires a significant amount of fresh water and generates a large quantity of slurry \cite{zhou2020role}. The government of Canada has reported that one barrel production of bitumen requires approximately about two to four barrels of water \cite{hussein2022recovery}. This leads the generation of a large amount of slurry, which is further transported to the tailing pond \cite{joshi2023estimation,zheng2024influence}. Thus, to transport the slurry, 
gravity driven open channels, particularly of rectangular and circular shape, are often used in the slurry disposal systems (SDS) \cite{fear2014conventional,kozicki1967non,kozicki1971improved,abulnaga2021slurry, spelay2007solids}. SDS aims to minimize land disturbance and water use. The content of fines significantly affects the rheological properties of the slurry mixture and results in the non-newtonian and viscoplastic behaviour of the mixture \cite{shook2002pipeline}. The fines in the slurry would increase its yield stress and apparent viscosity \cite{sumner2000rheology}. This variation of mixture fluid property makes the transportation of dewatered slurry non-economical under turbulent conditions \cite{chhabra1999non}, which could increase the power consumption in slurry transportation. However, an alternative approach is an open channel slurry transportation in the laminar flow conditions \cite{spelay2007solids}.


Over the years, researchers \cite{spelay2007solids,kozicki1967non,kozicki1971improved,matsuhisa1965analytical,haldenwang2004effect,haldenwang2010experimental,slatter2011laminar,javadi2015open} have attempted to study the flow behavior of non-Newtonian slurry transportation in open channels. Some of them \cite{kozicki1967non,kozicki1971improved,haldenwang2004effect} have performed the theoretical study and attempted to establish the empirical models for the estimation. However, the applicability of these models is limited because these were developed for homogeneous fluid and also with lack of experimental validation. Whereas some other researchers \cite{spelay2007solids,haldenwang2004effect,haldenwang2010experimental,slatter2011laminar,javadi2015open} have also experimentally evaluated the slurry flow behavior inside the open channels of different shapes. All studies were considered the uniform particle size to evaluate the flow behavior of non-newtonian thickened slurry inside the rectangular or semicircular flumes. However, in industrial applications, slurry consists of a wide range of PSD from fine (75µm) to coarse (700µm) \cite{sontti2023computational}. Change in particle kinetics due to variation of PSDs may affect the slurry flow behavior inside the flume, which ultimately leads to change in the settling behavior of solids \cite{feng2023flow}. This phenomenon might not be captured with the experimental study considering single particle size. Moreover, the effect of PSDs on the flow behavior of slurry inside a flume is not evaluated experimentally so far.\\

The computational fluid dynamic (CFD) is being progressively applied for the modelling of slurry transportation systems \cite{messa2021computational}. The long determination by researchers for this development is not just for eliminating the challenges faced during the experimental measurements \cite{sharma2023assessment,sharma2021measurement}, but also for the understanding and modelling of the complex mono-dispersed or poly-dispersed slurry flow phenomenon involved \cite{sontti2023computational}. \citet{spelay2007solids} have developed a code based on the visual basic for the modelling of non-newtonian slurry flow inside an open channel within the laminar regime and compared with the experimental data. However, the model applicability is limited to 1 dimensional (1D) slurry flow and is not reasonable to predict the solid settling phenomenon.\\


\noindent \citet{guang2009dns} investigated the turbulent non-newtonian fluid flow in the open channels using the direct numerical simulation (DNS) techniques. They modelled the shear-thining fluid inside the open channel with the HB fluid model and showed some fair agreement as well as disagreement with the experimental results. In addition, turbulent eddies were mentioned as influencing fluid flow behavior near channel bottom walls. Furthermore, recently \citet{sadeghi2023computational} and \citet{sontti2023computational} have modelled the poly-dispersed non-newtonian slurry flow in the long horizontal line. The developed models have shown reasonable validation with the available field measurements. \citet{sadeghi2023computational} determined that the PSD variation influences the fluid flow behavior as well as the bitumen recovery during the slurry transportation.\\

\noindent Bubbles in the flow have important implications for the wastewater treatment and oil water separation \cite{lohse2015surface,gao2021formation, zhang2011effects}. Microbubbles may improve the separation efficiency by flotation technology \cite{zhou2009role,zhang2011effects}.  \citet{sontti2023computational} stressed the effect of bubble size and volume fraction over the bitumen recovery in the industrial horizontal pipeline. They have shown the strong influence of bubbles on the velocity and concentration profile of solids inside the slurry transportation line. \\

In a recent work \citet{zhou2022microbubble, zhou2023enhanced} and \citet{motamed2020microbubble} experimentally determined the influence of microbubble inception over the bitumen recovery from the slurry stream. They have found that the bitumen recovery is increased to 50\% from a low to high range of solid content slurry transported through the pipeline. It has also been found by the researchers \cite{ma2023effect,jing2021settling, mokhtari2022effects} that the inception of bubbles of different volume fractions and size influence the settling behavior of solids inside the system. Thus, the presence of the bubble needs to be studied for the efficient recovery of bitumen with the significant settling of solids in the slurry transportation system. \\ 


Literature reveals that only a limited number of studies have been performed to investigate the non-newtonian slurry transportation in an open channel within the laminar regime. On the other hand, complex non-newtonian fluid behavior, PSDs and flume inclination play a significant role in the laminar flow of slurry and settling of solids inside the open channels. 
Moreover, the inception of bubbles in the slurry flow also causes the change in settling behavior of different range solid size\cite{jing2021settling}. The underlying phenomenon of bubble inception is desirable during the slurry transportation due to addition of water for the bitumen recovery in the extracted oil sand. To the best of the authors knowledge, limited studies are available which explored the laminar non-newtonian slurry transportation inside an open channel while considering the parameters like PSDs, channel inclination and bubble inception, which make it a complex multiphase flow system. Therefore, an in-depth understanding of slurry flow characteristics during open channel transportation is important to design an effective SDS along with significant recovery of water. \\

The main objective and novelty of this study is 3D unsteady E-E-based numerical modelling of the laminar non-newtonian slurry transportation inside a semicircular open channel. The general term ‘flume’ will be used throughout this paper to represent the semicircular open channel. The flow consists of poly-dispersed solids, non-newtonian carrier fluid and a small fraction of bubbles. The computational domain is considered from the experimental work of \citet{spelay2007solids}, and the numerical model is validated for the different range of available experimental parameters. A systematic parametric study is carried out to investigate the effect of parameters such as PSDs, flume inclination angle, bubble size and bubble volume fraction over the flow characteristics and settling behavior of multi-size solids slurry transportation through a flume. This fundamental understanding on the influence of these parameters may significantly benefit the industrial-scale dewatering process of the slurry disposal system. 

\section{Mathematical methodology } 
\subsection{Eulerian-Eulerian modelling } 
\noindent The present study employed the E-E modelling techniques to simulate the settling behavior of multisize particulate slurry inside the circular open channel. The E-E modelling approach provides the flexibility to simulate more secondary phases as interpenetrating continua \cite{parvathaneni2023effect,messa2021computational}. The coupling of different phases is performed by the interphase exchange coefficient and pressure. The governing equations are required to solve each phase separately.

Before formulating a system of equations describing the settling dynamics of particulate slurry, we considered a set of assumptions~\cite{fluent2022ansys}. Each phase is considered as a continuous medium, and momentum and continuity equations are solved for all phases. The phases can interpenetrate without any mass transfer between them. Surface tension effects are negligible, and there is no slip at phase interfaces.

Continuity equation
\begin{equation}
    \frac{\partial}{\partial t}\left(\alpha_p \rho_p\right)+\nabla \cdot\left(\alpha_p \rho_p \vec{v}_p\right)=0
    \label{eq:1}
\end{equation}
where $p$ shows the phases present in the system, which may be carrier fluid $f$, solid $s$, and bubble $b$.

Momentum equation\

for liquid phase;
\begin{align}
	\frac{\partial}{\partial t}\left(\alpha_f \rho_f \vec{v}_f\right) 
	+ \nabla \cdot \left(\alpha_f \rho_f \vec{v}_f \vec{v}_f\right)
	&= -\alpha_f \nabla P + \rho_f \alpha_f \vec{g} 
	+ \nabla \cdot \tau_f + \vec{F}_{f, s} +\vec{F}_{f, b}
	\label{eq:2a}
	\tag{2}
\end{align}

where $\vec{F}_{f, s}$ and $\vec{F}_{f, b}$ refer to the interface forces between liquid with the solid and bubble phases.\\

For solid phase;

\begin{align}
  \frac{\partial}{\partial t}\left(\alpha_{si} \rho_{si} \vec{v}_{\mathrm{si}}\right) 
  &+ \nabla \cdot \left(\alpha_{\mathrm{si}} \rho_{si} \vec{v}_{si} \otimes \vec{v}_{si}\right)
  = -\alpha_{si} \nabla P - \nabla P_{si} + \rho_f \alpha_{si} \vec{g} \nonumber \\
  &\quad + \nabla \cdot \boldsymbol{\tau}_{si} 
  + \vec{F}_{si, f} 
  + \vec{F}_{\text{drag, s, b}} 
  + \beta_{ij}\left(\vec{v}_{sj} - \vec{v}_{si}\right)
  \tag{3} \label{eq:2b}
\end{align}

where, $\nabla P$ is the static pressure gradient, $\rho_f \alpha \mathbf{g}$ is the body force, $\nabla P_{si}$ is the solid pressure gradient and
\begin{align}
  \beta_{i j}=\frac{3\left(1+e_{i j}\right)\left(\frac{\pi}{2}+{C}_{{ft}, i j} \frac{\pi^2}{8}\right) \alpha_{s i} \rho_{s i} \alpha_{s j}\left(d_{s i}+d_{s j}\right)^2 g_{0, \mathrm{ij}}}{2 \pi\left(\rho_{s i} d_{s i}^3+\rho_{s j} d_{s j}^3\right)}  \label{eq:2c}
    \tag{4}
\end{align}
\begin{align}
\tau_{s i}=\alpha_{s i} \mu_{s i}\left(\nabla \overrightarrow{{v}}_{s i}+\left(\nabla \overrightarrow{{v}}_{s i}\right)^{{T}}\right)+\alpha_{s i}\left(\lambda_{s i}-\frac{2}{3} \mu_{s i}\right)\left(\nabla \cdot \overrightarrow{{v}}_{s i}\right) \overline{\overline{\mathbf{I}}}  \label{eq:2d}
    \tag{5}
\end{align}
\begin{align}
\lambda_{s i}=\frac{4}{3} \alpha^2 \rho_{s i} d_{s i} \vec g_{0, i i}\left(1+e_{i j}\right)\left(\frac{\Theta_{s i}}{\pi}\right)^{1 / 2}  \label{eq:2e}
    \tag{6}
\end{align}
\begin{align}
\vec{F}_{\text {drag }, s}=-\vec{F}_{\text {drag }, f}  \label{eq:2f}
    \tag{7}
\end{align}
\begin{align}
\vec{F}_{d t, s}=-\vec{F}_{t d, f} \label{eq:2g}
    \tag{8}
\end{align}
\begin{align}
\frac{\partial}{\partial t}\left(\alpha_b \rho_b \vec{v}_b\right)+\nabla \cdot\left(\alpha_b \rho_b \vec{v}_b \otimes \vec{v}_b\right)=-\alpha_b \nabla P+\rho_f \alpha_b \vec{g}+\nabla \cdot \mu_b\left(\nabla \vec{v}+\nabla \vec{v}^T\right)+\vec{F}_{b, f}+\vec{F}_{b, s i} \label{eq:3a}
    \tag{9}
\end{align}
\begin{align}
    \vec{F}_{\mathrm{b}, \mathrm{f}}=-\vec{F}_{f, b} ; \vec{F}_{\mathrm{b}, \mathrm{si}}=-\vec{F}_{s i, b} \label{eq:3b}
    \tag{10}
\end{align}
The equations are used to establish the mass and momentum exchange between the primary and secondary phases.


Particle-particle interaction plays an important role in the flow behavior of the particulate system. Thus, to capture this interaction, the Kinetic theory of granular flow (KTGF) is applied in this study. For each granular (solid) phase, the pressure is calculated using the relation given by Gidaspow \cite{gidaspow1994multiphase}.

\begin{align}
 P_{s i}=\alpha_{s i} \rho_{s i} \Theta_{s i}\left[1+2 \sum_{j=1}^2\left(\frac{d_{s i}+d_{s j}}{2 d_{s i}}\right)^3\left(1+e_{i j}\right) \alpha_{s j} \vec g_{0, i j}\right] \label{eq:4a}
    \tag{11}
\end{align}

where $d_{s i}$ and $d_{s j}$ are the $i^{th}$ and $j^{th}$ phase particle diameters and $\vec g_{0, i j}$  is the radial distribution function for the solid phase estimated as follows \cite{lun1984kinetic}; 

\begin{align}
\vec g_{0, \mathrm{ii}}=\left[1-\left(\sum_{\mathrm{i}=1}^2 \alpha_{s i} / \alpha_{s, \max }\right)^{1 / 3}\right]^{-1}+\frac{d_{s i}}{2} \sum_{\mathrm{i}=1}^2 \frac{\alpha_{s i}}{d_{s i}} \label{eq:4b}
    \tag{12}
\end{align}
$\Theta_{si}$ is the granular temperature for solid phase and is calculated as: 
\begin{align}
\Theta_{s i}=\frac{1}{3}\left\|\vec{v}_{s i}^{\prime}\right\|^2 \label{eq:4c}
    \tag{13}
\end{align}

Further, Ding and Gidaspow \cite{fluent2022ansys} reported that the granular temperature for the solid phase can be estimated as follows:
\begin{align}
0=\left(-P_{s i} \overline{\bar{I}}+\overline{\bar{\tau}}_{s i}\right): \nabla \vec{v}_{s i}-\gamma_{\Theta_{s i}}+\phi_{f i} \label{eq:4d}
    \tag{14}
\end{align}

where $\left(-P_{s i} \overline{\bar{I}}+\overline{\bar{\tau}}_{s i}\right)$ is the formation of the fluctuation energy due to shear in the solid phase, $\Theta_{si}$ is the energy dissipation rate due to collision between solid particles and estimated as follows \cite{lun1984kinetic,parvathaneni2021eulerian}; 
\begin{align}
\gamma_{\Theta_{si}}=\frac{12\left(1-e_{i i}^2\right) \vec g_{0,i i}}{d_{s i} \pi^{1 / 2}} \rho_{s i} \alpha_{s i}^2 \Theta_i^{3 / 2} \label{eq:4e}
    \tag{15}
\end{align}

$\phi_{f i}$ used to show the transfer of kinetic energy due to random change in the velocity of the granular phase to the fluid phase and can be calculated by \cite{fluent2022ansys};
\begin{align}
\phi_{f i}=-3 K_{f i} \Theta_i \label{eq:example-a}
    \tag{15}
\end{align}

Further, the coupling of bulk viscosity of solid phase for solving the momentum equation is performed by putting the solid viscosity in the shear-stress tensor term as given in equations 5 and 6. The shear viscosity $(\mu_{si}) $of the granular phase is defined as follows; 
\begin{align}
\mu_{s i}=\mu_{s i, \mathrm{col}}+\mu_{s i, \mathrm{kin}}+\mu_{s i, \mathrm{fr}}\label{eq:example-a}
    \tag{16}
\end{align}

Where three shear viscosities, collisional viscosity $\left(\mu_{si,col}\right)$ \cite{syamlal1993mfix}, kinetic viscosity $\left(\mu_{si,kkin }\right) \cite{gidaspow1991hydrodynamics}$ and frictional viscosity $\left(\mu_{si, fr.}\right) \cite{schaeffer1987instability,parvathaneni2023eulerian}$ are as follows;
\begin{align}
\mu_{s i, \mathrm{col}}=\frac{4}{5} \alpha_{s i} \rho_s d_{s i} \vec g_{0, i i}\left(1+e_{i j}\right)\left(\frac{\Theta_{s i}}{\pi}\right)^{1 / 2} \alpha_{s i} \label{eq:example-a}
    \tag{17}
\end{align}

\begin{align}
\mu_{s i, \mathrm{kin}}=\frac{10 \rho_{s i} d_{s i}\left(\Theta_{s i} \pi\right)^{1 / 2}}{96 \alpha_{s i}\left(1+\mathrm{e}_{i j}\right) \vec {g}_{0, \mathrm{ii}}}\left[1+\frac{4}{5} \mathrm{~g}_{0, \mathrm{ii}} \alpha_{s i}\left(1+\mathrm{e}_{i i}\right)\right]^2 \alpha_{s i} \label{eq:example-a}
    \tag{18}
\end{align}
\begin{align}
\mu_{s i, \mathrm{fr}}=\frac{P_{s i} \sin \varphi_{s i}}{2 I_{2 \mathrm{D}}^{1 / 2}} \label{eq:example-a}
    \tag{19}
\end{align}
All reported equations are used to model the granular flow. A nomenclature section is also reported to explain the symbols used in the governing equations.\

\subsection{Rheological fluid model}

\citet{visintainer2023slurry} mentioned that the non-newtonian slurry, which contains medium to a high-range concentration of fines, especially where clay-type particles are involved in the mixture, shows the behavior of Bingham, Casson and HB fluid models. The HB model is able to represent a wide range of slurry, however, it requires extra fitting curve parameters. The HB shows the power-law fluid with yield stress, and the equation for the model will be as follows;
\begin{align}
\tau=\tau_y+k \dot{\gamma}^n \label{eq:example-a}
    \tag{20}
\end{align}
where \textit{$\tau$} is the shear stress, $\tau_y$ is the yield stress, $k$ is the consistency index, $\dot{\gamma}$ is the shear rate, and $n$ is the flow index. In the present study, we have considered the medium-range concentration of fines as were used by \citet{spelay2007solids}. The detailed properties of the carrier fluid are presented in Table \ref{tab:FluidProperties}. Moreover, through the experimental study, \citet{spelay2007solids} reported that the same fine content carrier fluid behaves like the Bingham fluid with a yield stress, $\tau_{\mathrm{y}}$ of 40 Pa and consistency index, $k$ as 0.04 Pa.s, as mentioned in Table \ref{tab:FluidProperties}. For this instance, in our present study, we have taken the carrier fluid as Bingham fluid. Therefore, to model the Bingham fluid, we have used the flow index $n$=1 in the HB fluid model\cite{moreno2016modelling} and performed modelling using the HB model in the ANSYS fluent software for the Laminar flow condition.

\subsection{Computational domain}
\begin{figure}
	\centering
	\includegraphics[width=\textwidth]{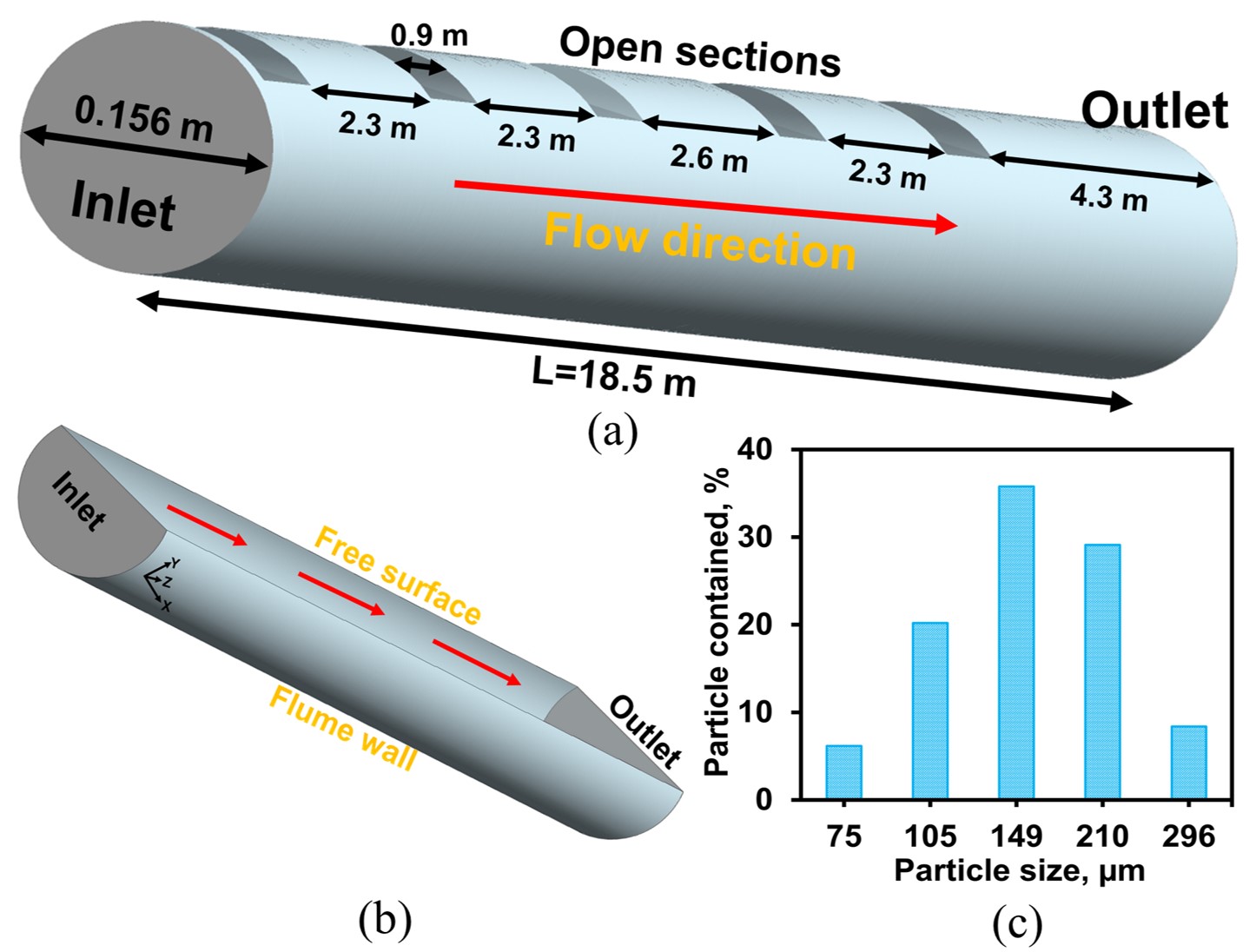}
	\caption{ (a) Slurry transportation flume with dimensions, (b) simplified computational domain (c) particle size distribution\cite{spelay2007solids}.}
 \label{fig:Model1}
\end{figure}

The present study considers a three-dimensional (3D) computational domain of the open channel fume section for numerical investigation. Figure \ref{fig:Model1}(a) shows the slurry transportation flume with dimensions which was used by \citet{spelay2007solids} for the experimental measurement. The transportation line is about 18.5 m long and 0.156 m in diameter. It can be seen from Figure \ref{fig:Model1}(a) that the line is opened at five different sections at a definite distance. The realistic system should be modelled by using the entrainment of air-water-solid simultaneously. However, due to challenges in the modelling of three different Eulerian phases \cite{guo2021analysis}, the computational domain is modified and set up for the constant depth of flow for each operating condition. For this instance the domain is simplified as shown in Figure \ref{fig:Model1}(b). It shows the 3D simplified computational domain which is generated by truncating the flume to the height of constant depth of the flow. The domain is developed using the ICEM CFD tool. The modified domain consists of an inlet, pipe wall, outlet and free surface over the full length of the flume.               

\subsection{Modelling strategy} 

\subsubsection{Particles in the slurry}
The present numerical investigation is performed by considering the experimental program given in \citet{spelay2007solids}. The multisize particulate slurry is used to investigate the flow behavior of thickened slurry with and without the effect of the bubbles. \citet{spelay2007solids} mentioned that the dry sieve analysis determined the particle size distribution and found the mean particle diameter in the range 75-296 $\mu$m with a median diameter of $d_{50}$ -188 $\mu$m. Figure \ref{fig:Model1}(c) shows the distribution of mean particle size of unimin sand, which is employed in the flume for the experimental investigation. \citet{schaan2000effect} has mentioned that the maximum packing factor is independent of particle geometry. Thus, through experimentation \citet{spelay2007solids} has found that the average packing limit for the current set of sand samples is about 58.2\%.
 

The thickened slurry used in the present study consists of fine kaolin clay particles, unimin sand particles and water. The mixing of fine clay particles and water results in a non-newtonian fluid which works as a carrier for solid transportation in the current system. The physical properties of the carrier fluid and sand particles are given in Table \ref{tab:FluidProperties}.

\begin{table}[!ht]
    \centering
    \caption{Physical property of the slurry. \cite{spelay2007solids}}
    \label{tab:FluidProperties}
    \small 
    \setlength{\tabcolsep}{40pt} 
    \begin{tabular}{cc}
        \hline
        \hline
        \textbf{Parameter} & \textbf{Value} \\
        \hline
        \hline
        Carrier fluid composition & Fine kaolin clay, silica sand and water\\
        Carrier fluid density, $\rho_{\mathrm{f}}$ (kg/m$^3$) & 1303 \\
        Sand density, $\rho_{\mathrm{S}}$ (kg/m$^3$) & 2650 \\
        Yield stress, $\tau_{\mathrm{y}}$ (Pa) & 40 \\
        Power-law index, $n$ & 1 \\
        Constant, ${k}$ (Pa.s) & 0.04 \\
        Mixture density, $\rho_{\mathrm{m}}$ (kg/m$^3$) & 1510 \\
        \hline
        \hline
    \end{tabular}%
\end{table}

\subsubsection{Boundary conditions and solution methodology}
\begin{figure}
	\centering
\includegraphics[width=\textwidth]{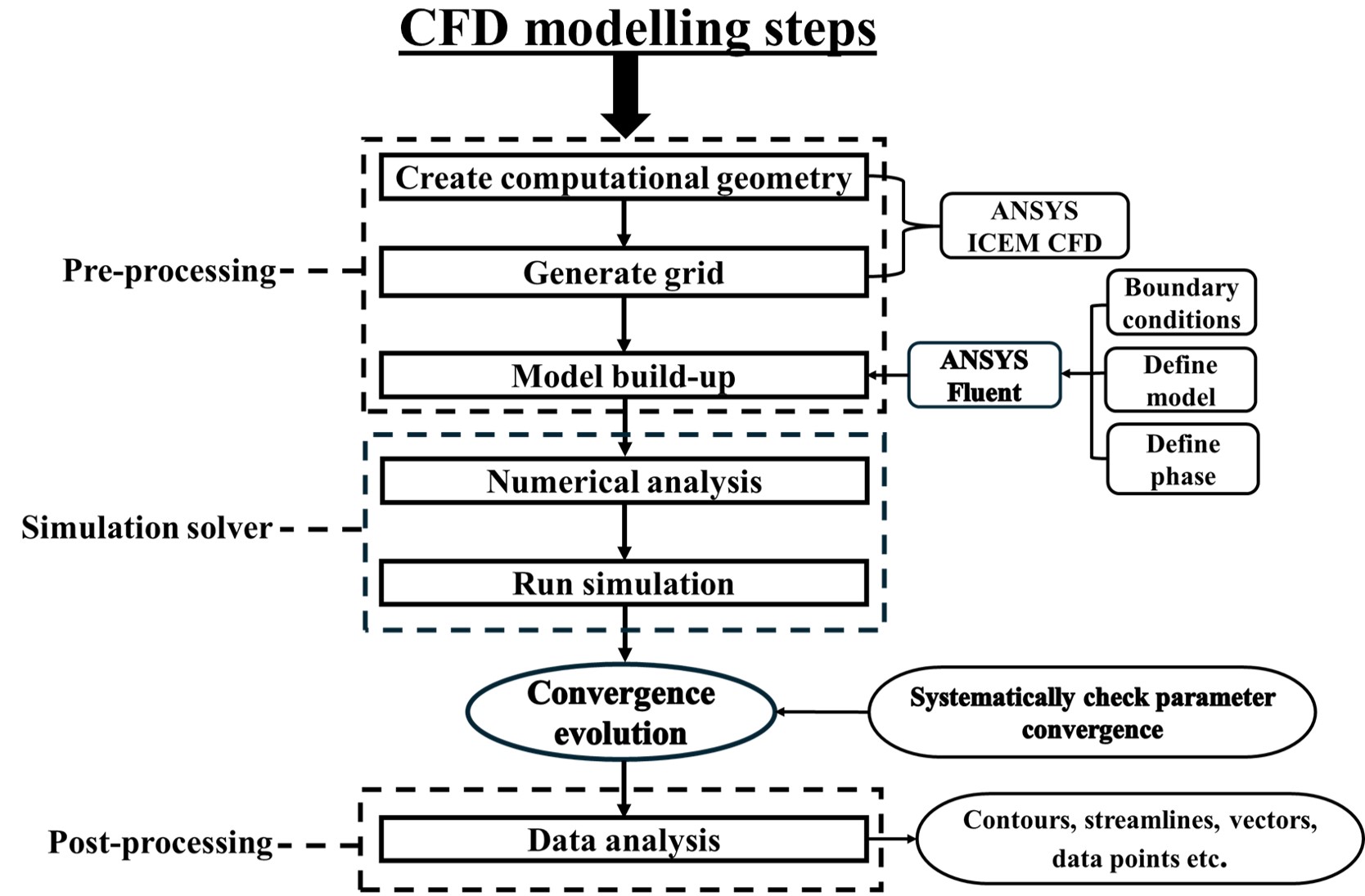}
\caption{Flow chart for the modelling steps.}
\label{fig:Model2}
\end{figure}

\vspace{-0.3cm}
The numerical modelling is performed using the commercial software \textit{ANSYS} Fluent Solver 2020 R2, and it works on the finite volume method. All simulations used the unsteady Navier-Stokes (NS) equations for solving the motion and interacting forces in the granular flow \cite{fluent2022ansys}. Figure \ref{fig:Model2} presented the complete steps of the numerical modelling which is used in the present study. The E-E model with poly-dispersed phases is considered for the simulation in the present study. Total of seven Eulerain phases are considered for the modelling, which consists of carrier fluid, solid and bubble. Carrier fluid is considered as a primary phase and the five different size sand particles and bubble are considered as secondary phases. Further using the drag model implementation, the intersection of each phase has been established in the simulation. The bulk velocity and solid volume fraction of each phase are defined at the inlet of the flume. The top surface is open to the atmosphere; thus, no shear boundary is specified for the open channel condition. The flume wall is chosen as a no-slip boundary condition, and for the outlet atmospheric pressure condition is imposed.   Table
\ref{tab:boundary_and_solution_params} reported the detailed boundary condition and solution parameters. \\

\begin{table}[!ht]
    \centering
    \caption{Boundary condition and solution parameters.}
    \label{tab:boundary_and_solution_params}
    \renewcommand{\arraystretch}{0.8} 
    \setlength{\tabcolsep}{12pt} 
    \begin{tabular}[t]{@{}cp{0.5\textwidth}@{}}
        \hline
        \hline
        \textbf{Parameters} & \textbf{Modelling} \\
        \hline
        \hline
        Multiphase model & Eulerian-Eulerian \\
        Viscous regime & Laminar \\
        Primary phase & Carrier fluid \\
        Secondary phases & Sand, bubble \\
        Number of phases & Seven: carrier fluid, five solids, bubble \\
        Boundary conditions & Inlet: velocity, volume fraction \\
        & Outlet: pressure outlet \\
        & Free surface: no shear \\
        & Flume wall: no slip \\
        Simulation type & Transient: first-order implicit \\
        & Time step: 0.001 \\
        & Number of time steps: 21,000 \\
        Discretization scheme & Pressure: PRESTO \\
        & Momentum: second order upwind \\
        & Volume fraction: first order upwind \\
        Under relaxation factor & Pressure: 0.3 \\
        & Momentum equation: 0.7 \\
        & Volume fraction: 0.4 \\
        Pressure-velocity coupling & Phase coupled SIMPLE \\
        Carrier viscosity & HB model \cite{visintainer2023slurry}  \\
        Carrier-solid drag & Symmetric \\
        Carrier-bubble drag & Symmetric \\
        Solid-solid interaction & Symmetric drag model\\
        Solid-bubble interaction & Symmetric drag model \\
        Particle restitution coefficient & 0.9 \cite{sontti2023computational}\\ 
        Convergence criteria & $\leq 1.0 \;\;\text{E}-4$ \\ \\
        \hline
        \hline
    \end{tabular}
\end{table}
The selection of desired time steps for the simulation, grid independency test, drag model selection and model validation would be carefully considered for the numerical modelling and investigation. Thus, the preliminary first set of study for the grid and drag model selection is performed for the operating condition of case 1, as mentioned in Table \ref{tab:experimental-parameters}, with the particle size as median particle diameter ($d_{50}$~=~188~$\mu$m). Table~\ref{tab:experimental-parameters} presents the experimental condition from the \citet{spelay2007solids} used as an input parameter for the simulation of model validation. \

\begin{table}[!ht]
    \centering
    \caption{Experimental parameters for numerical model validation. \cite{spelay2007solids}}
    \label{tab:experimental-parameters}
    \resizebox{1\textwidth}{!}{%
        \begin{tabular}{cccccccc}
            \hline
            \hline
            Case & \makecell{Slurry \\ type} & \makecell{Flow \\ rate  (L/s)} & \makecell{Bulk\\ velocity  (m/s)} & \makecell{Flume \\ inclination \\ angle (degrees)} & \makecell{Sand particle \\ size, $d_{50}$ ($\mu$m)} & \makecell{Depth of \\ flow  (m)} & \makecell{Solid volume \\ fraction\\ } \\
            \hline
            \hline
            1 & \makecell{Thickened \\ slurry} & 5 & 0.41 & 4 & 188 & 0.1039 & 0.125 \\
            2 & \makecell{Thickened \\ slurry} & 2.5 & 0.19 & 4.5 & 188 & 0.1077 & 0.10 \\
            3 & \makecell{Thickened \\ slurry} & 5 & 0.44 & 4.5 & 188 & 0.0968 & 0.11 \\
            4 & \makecell{Thickened \\ slurry} & 5 & 0.5 & 5.4 & 188 & 0.0861 & 0.11 \\
            \hline
            \hline
        \end{tabular}%
    }
\end{table}

\subsection{Grid selection}

\begin{figure}
	\centering
\includegraphics[width=0.8\textwidth]{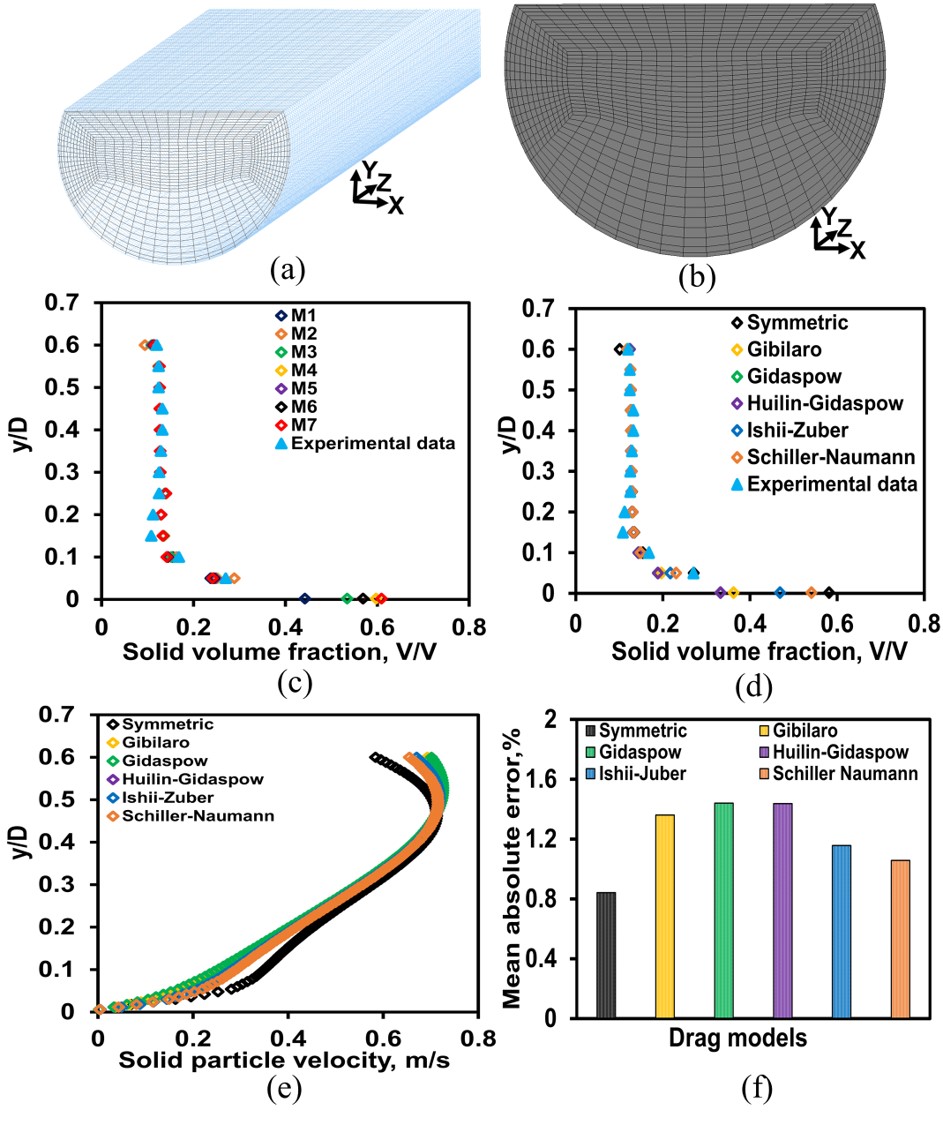} \hfill
\caption{(a) Grid structure along the length and (b) cross-section of the flume (c) variation of solid distribution for different sets of grids, and comparison of drag models (d) chord average solid volume fraction (e)velocity distribution and (f) mean absolute error of solid volume fraction over the plane at 14.5 m from the inlet of the flume along the length.}
\label{fig:Model3}
\end{figure}

The optimized grid for the simplified computational domain is required for the study. Thus, seven different extra coarse grids M1 to extra fine grid M7 have been developed and used for grid independence test. Table \ref{tab:number_of_elements} presents the mesh parameters for all sets of grids. The number of elements varied from 0.24 million to 1.03 million for M1 to M7, respectively. Figures \ref{fig:Model3}(a,b) show the mesh structure along the length and cross-section of the flume respectively. The hexahedral O grid mesh with 0.001~$m$ thickness from the wall is developed for the specified computational domain, which shows minimum orthogonal as 0.64 and aspect ratio $<$51.

\begin{table}[ht]
    \centering
    \caption{Number of control volumes (CV) for different meshes.}
    \label{tab:number_of_elements}
    \begin{tabularx}{\textwidth}{c *{7}{>{\centering\arraybackslash}X}}
        \hline
        \hline
        \multirow{2}{*}{Parameters} & \multicolumn{7}{c}{Mesh} \\
        & M1 & M2 & M3 & M4 & M5 & M6 & M7 \\
        \hline
        \makecell{Number \\ of elements} & 242,995 & 357,603 & 436,947 & 570,840 & 683,795 & 833,667 & 1,036,435 \\
        \makecell{ Uncertainty in predicted \\ average solid velocity \\ \%} & 11.43 & 9.68 & 4.51 & 4.72 & 0.78 & 1.60 & 1.04 \\
        \hline
        \hline
    \end{tabularx}
\end{table}

The grid independence test is performed to keep the balance between the computation time and the accuracy of the simulation. For this purpose, the simulation is performed with all seven sets of grids for the slurry flow under the operating condition of case 1. The predicted chord average solid volume fraction and centre line solid particle velocity over the plane at 14.5~m from the inlet of the flume are used for the validation with the experimental data of \citet{spelay2007solids}. Figure \ref{fig:Model3}(c) shows the predicted chord average solid volume fraction from the different sets of grids and the experimental data. It is observed that the particle settling distribution from the top free surface to the flume wall is in-line with the experimental data for all sets of grids. However, in the zone $y/D$$<$0.3 the value solid volume fraction is closer to experimental data for the grids M5 to M7 than M1 to M4. Thus, the predicted average sand velocity over the same plane is also compared with the experimental bulk velocity of the fluid. Table \ref{tab:number_of_elements} shows the uncertainty percentage in the predicted average velocity from the bulk velocity over the plane at 14.5 m from the inlet of the flume.  The deviation of the predicted value for the grid M5 is the least among all i.e. about 0.78\% and is below than 1\% which is in the acceptable range\cite{zheng2024influence}. Also, the predicted velocity from grid M5 shows about 0.80\% and 0.20\% less uncertainty from grid M6 and M7, respectively. Therefore, M5 grid of 0.68 million hexahedral elements is selected for the further numerical investigation.

\subsection{Drag model selection}
The interaction between two Eulerian phases could be predicted by drag models in the CFD simulation of multiphase systems. \citet{liu2022effect} reported that the E-E modelling is an efficient technique for simulating solid-liquid slurry systems, but its reliability depends upon the accurate defining of the liquid-particle interaction during the CFD simulation. Moreover, \citet{spelay2007solids} mentioned that in the multiparticle system, which consists of solids greater than 5-10\% by volume, the interaction of neighbouring particles influences the settling behavior of slurry solids. For this instance, the sensitivity of the drag models for the current set of operating conditions needs to be investigated. Literature reveals that the various established drag models like Symmetric \cite{fluent2022ansys}, Gibilaro \cite{gibilaro1985generalized}, Gidaspow \cite{gidaspow1994one}, Huillin-Gidaspow \cite{huilin2003hydrodynamics}, Ishii \cite{Ishii1987TWOFLUIDMF}, Schiller-Naumann \cite{schiller1933drag} are available with the researchers.\\


Figure \ref{fig:Model3}(d) compares predicted chord average solid volume fraction data with experimental measurements \cite{spelay2007solids} for different carrier-solid drag models. It can be seen that all selected models have shown a similar trend with experimental data. Figure \ref{fig:Model4}(e) presents the variation of the solid particle velocity inside the flume. The examination of solid particle velocity shows the decreasing trend from the free surface to bottom of the flume for all drag models. However, a slight difference is observed in the predicted solid particle velocity for the top and bottom of the flow, but the trend is in line with the \citet{spelay2007solids} experiment observations. Thus, to compare the accuracy of different drag models, a quantitative analysis of solid distribution using mean absolute error (MAE) is also performed. The percentage of MAE is defined as follows \cite{javadi2015laminar};

\begin{align}
\% \text { of } \mathrm{MAE}=\left(\frac{1}{\mathrm{~N}} \sum_{\mathrm{i}=1}^{\mathrm{N}}\left|\mathrm{p}_{\mathrm{i}}-\mathrm{m}_{\mathrm{i}}\right|\right) \times 100 \label{eq:example-a}
    \tag{21}
\end{align}

Where $p_{i}$ and $m_{i}$ are the predicted and measured values of the solid volume fraction, respectively. N is used to represent the number of measurements. MAE measures the average error made in the predicted value of solid volume fraction with each drag model. Figure \ref{fig:Model3}(f) presents the percentage of error observed for the solid volume fraction predicted by different drag models. It indicates that the lowest error in predicted solid volume fraction is obtained by the symmetric drag model. Thus, in further all simulation steps, the symmetric drag model is considered to establish the interaction between carrier fluid and solid. Furthermore, the interaction between particle-particle and particle-bubble is also estimated using the symmetric model \cite{sadeghi2023computational,sontti2023computational}.\

\subsection{Parameters for numerical investigation}

Six sets of numerical simulations of the open channel thickened slurry flow are performed with multi-size particulate slurry at different operating conditions. All cases have been run for the non-newtonian slurry consisting of fine kaolin clay, solids and water with different operating conditions. The first set of simulations was performed for the grid and drag model selection at the operating condition for case 1. The second set of simulations is performed for the other four different operating conditions as mentioned in Table \ref{tab:experimental-parameters}. 
\begin{table}[!ht]
    \centering
    \caption{Particle size distribution in polydispersed slurry sample.}
    \label{tab:Particlesize}
    \begin{tabular}{c@{\hspace{8mm}}c@{\hspace{8mm}}c@{\hspace{8mm}}c@{\hspace{8mm}}c@{\hspace{8mm}}c}
        \hline
        \hline
        \multirow{2}{*}{Sample} & \multicolumn{5}{c}{Mean particle size $(\mu \mathrm{m})$} \\
        \cline{2-6}
        number & Solid 1-75 & Solid 2-105 & Solid 3-149 & Solid 4-210 & Solid 5-296 \\
        \cline{2-6}
        & \multicolumn{5}{c}{\% of volume fraction} \\
        \cline{2-6}
        PSD 1 & 6.20 & 20.30 & 35.90 & 29.21 & 8.39 \\
        PSD 2 & 20.00 & 20.00 & 20.00 & 20.00 & 20.00 \\
        PSD 3 & 6.20 & 8.39 & 20.30 & 29.21 & 35.90 \\
        PSD 4 & 35.90 & 29.21 & 20.30 & 8.39 & 6.20 \\
        \hline
        \hline
    \end{tabular}
\end{table}

\begin{table}[!ht]
    \centering
    \caption{Simulation parameters configuration. }
    \label{tab:operating_parameters}
    \begin{tabularx}{\textwidth}{@{}c *{5}{X}@{}}
        \hline
        \hline
        & \multicolumn{5}{c}{Operating conditions} \\
        \cline{2-6}
        Parameters & Cases & PSD & \begin{tabular}{c} Bubble \\ size \\ $(\mu \mathrm{m})$ \end{tabular} & \begin{tabular}{c} Bubble volume \\ fraction \end{tabular} & \begin{tabular}{c} Inclination \\ angle \\ (degree) \end{tabular} \\
        PSD & Case 1 & \begin{tabular}{@{}c@{}} PSD 1, PSD 2, \\ PSD 3, PSD 4 \end{tabular} & - & - & 4 \\
        Flume \\ inclination & Case 1 & PSD 1 & - & - & $3,4,5,6$ \\
        \begin{tabular}{@{}c@{}} Bubble \\ size \end{tabular} & Case 1 & PSD 1 & \begin{tabular}{@{}c@{}} $5,50,500,$ \\ $1000$ \end{tabular} & 0.0025 & 4 \\
        \begin{tabular}{@{}c@{}} Bubble volume \\ fraction \end{tabular} & Case 1 & PSD 1 &  50 & \begin{tabular}{@{}c@{}} 0.0025, 0.01 \\ 0.02, 0.03 \end{tabular}  &  4 \\
        \hline
        \hline
    \end{tabularx}
\end{table}
Further, the remaining set of simulations were performed for the parametric study. The effect of particle size distribution (PSD) is investigated by keeping the total solid concentration constant and varying the particle size volume fraction. Table \ref{tab:Particlesize} depicts the PSD information of the four different cases. All the samples are prepared by considering five different sizes of particles as reported in the Figure \ref{fig:Model1}(c). PSD 1 represents the base case for the parametric study, which is same as the experimental study. PSD 2 represents the equal percentage of volume fraction for all size of particles. The arrangement of solid volume fractions of the solids in increasing and decreasing order resulted in PSD 3 and PSD 4, respectively.\

Further, the simulations are performed for different sets of flume inclination angles, bubble size and bubble volume fraction. Table \ref{tab:operating_parameters} reports the operating condition at which the parametric study is performed and discussed. \

\subsection{Model validation}

\begin{figure}
	\centering
\includegraphics[width=1\textwidth]{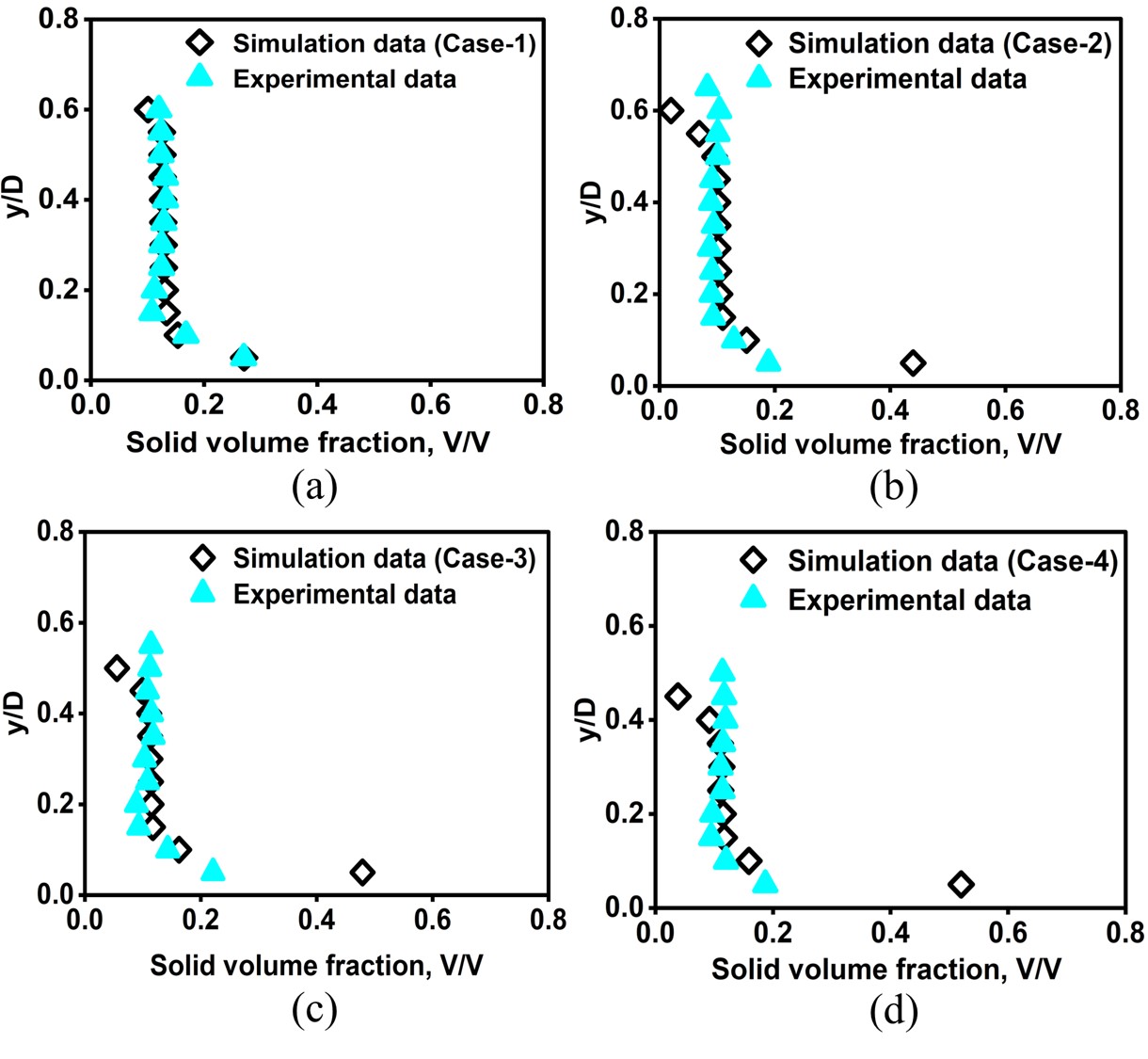}
\caption{Comparison of experimental \cite{spelay2007solids} and predicted solid volume fraction data for the different cases at a median diameter of 188 $\mu$m (a) Case 1, (b) Case 2, (c) Case 3, and (d) Case 4 over the plane at 14.5 m from the inlet along the length.}
\label{fig:Model4}
\end{figure}

\begin{figure}[!ht]
	\centering
\includegraphics[width=1\textwidth]{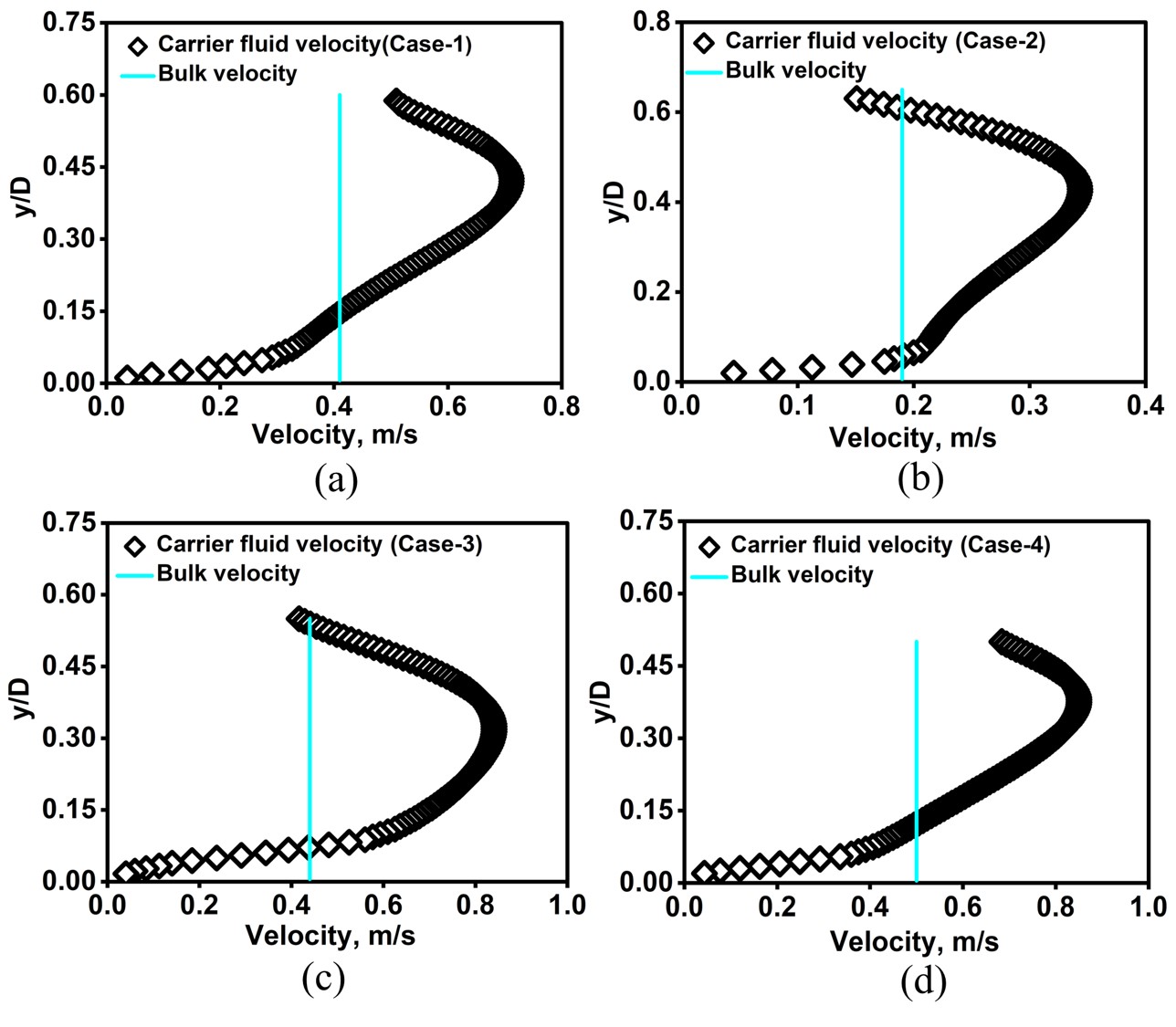}
\caption{Comparison of experimental and simulation data for the different cases at a median diameter of 188µm (a) Case 1 (b) Case 2 (c) Case 3, and (d) Case 4 over the plane at 14.5 m from the inlet along the length.}
\label{fig:Model6}
\end{figure}

The accuracy and validity of the developed numerical model are verified by experimental data as given in \citet{spelay2007solids}. The solid volume fraction and velocity trend at different operating conditions are compared and presented in Figures \ref{fig:Model4}(a-d) and \ref{fig:Model6}(a-d). All results are presented as a function of vertical distance and normalized by the diameter of the circular section.\

\begin{figure}[!ht]
	\centering
\includegraphics[width=0.7\textwidth]{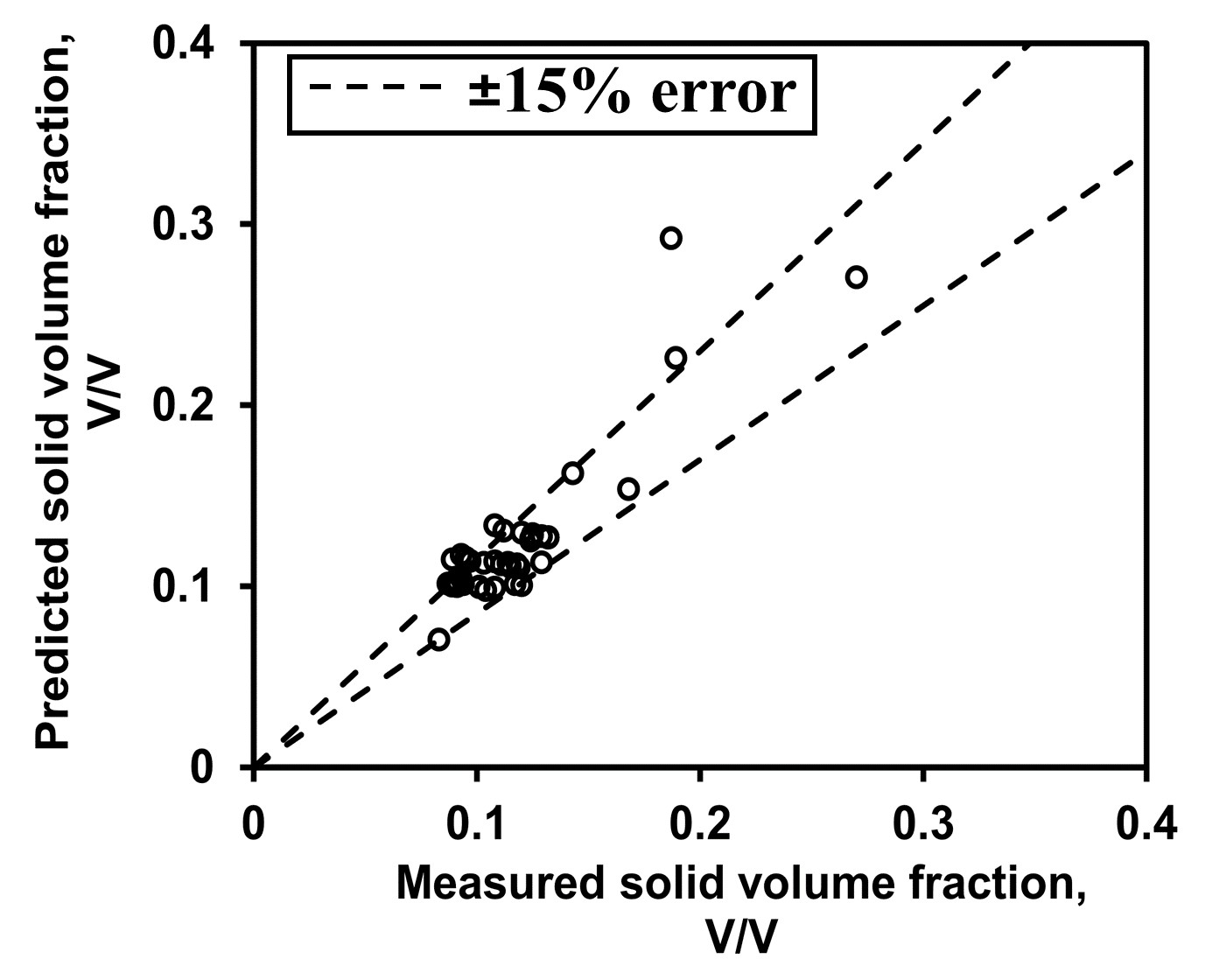}
\caption{Uncertainty band representation to compare the measured and predicted solid volume fraction. }
\label{fig:Model5}
\end{figure}

Figures \ref{fig:Model4}(a-d) show the measured and predicted chord average solid volume fraction at the plane 14.5 m from the inlet for different cases, as in Table \ref{tab:experimental-parameters}. The solid particle distribution is found to be asymmetric, with a larger concentration in the bottom of the flume at around $y/D$ $<0.20$. 
However, an acceptable trend agreement with the experimental data is obtained with a slight over-prediction of particle concentration at the bottom of the flume. Figure \ref{fig:Model5} depicts the uncertainty band to compare the measured and predicted value of solid volume at different positions over a plane for all cases. It is found that the model predicts the solid volume fraction within the ±15\% uncertainty band for approximately 93\% of data.\\

\begin{figure}
	\centering
\includegraphics[width=0.7\textwidth]{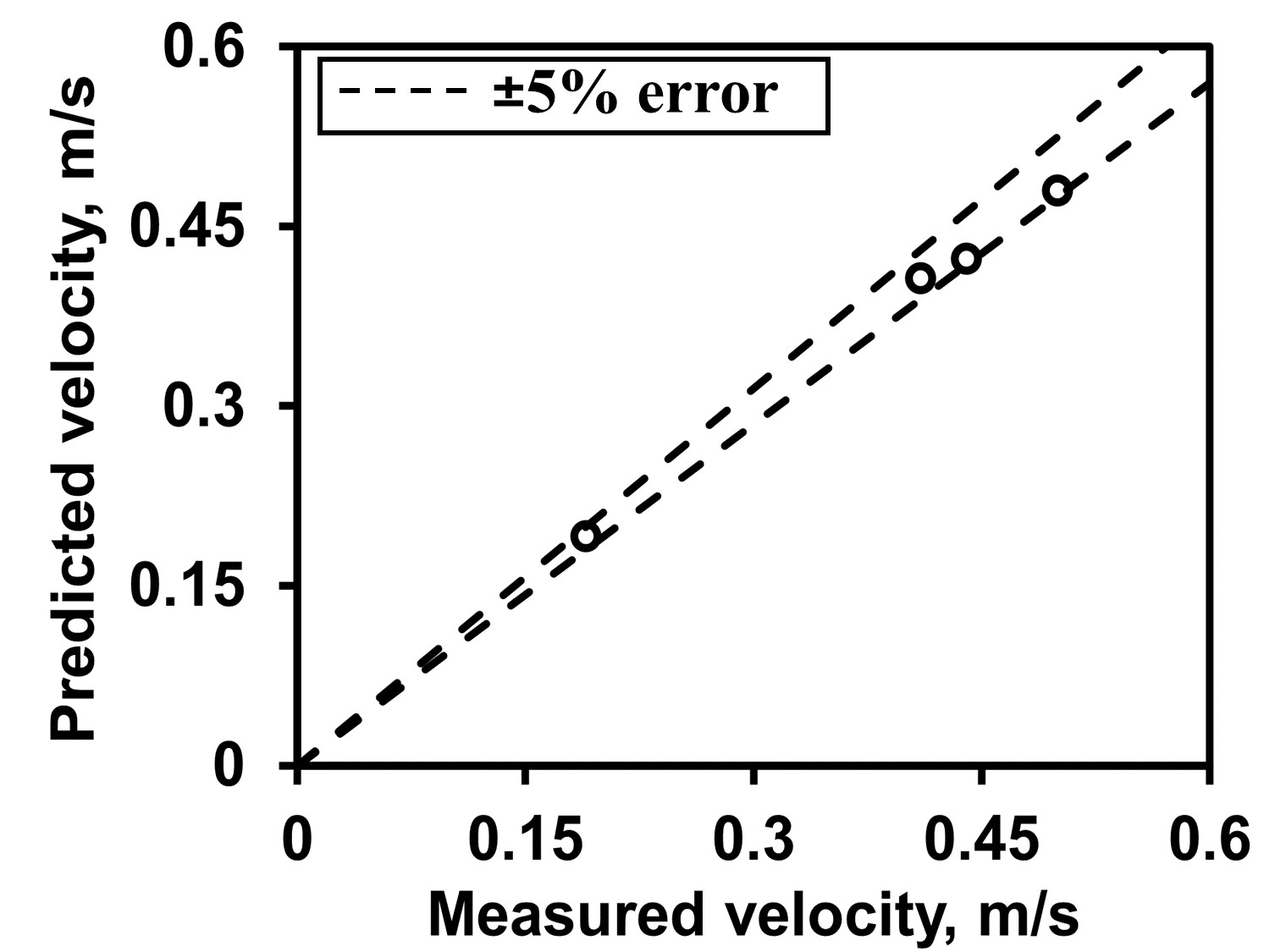}
\caption{Uncertainty band representation to compare the measured bulk flow velocity and predicted average flow velocity of carrier fluid over the plane at 14.5 m from the inlet along the length.}
\label{fig:Model7}
\end{figure}

Figures \ref{fig:Model6}(a-d) present the vertical centre line velocity of carrier fluid for different cases over the plane 14.5 m from the inlet. The carrier fluid velocity increases along the centre line of the pipe as one travels from the bottom of the flume to the free surface. The maximum velocity is always found to be less than twice of bulk flow velocity, which shows that the carrier fluid behaves like the shear-thinning fluid in the flume \cite{eesa2009cfd}. Moreover, the trend of velocity is most accurately represented by the bingham fluid model \cite{shook2002pipeline}. Figure \ref{fig:Model7} compares the predicted average velocity over a plane at 14.5 m to the measured bulk velocity. It is found that for all cases the uncertainty margin is about ±5\%, which can be considered a reliable model case.  \

\section{Results and discussion}
\begin{figure}[!ht]
	\centering
\includegraphics[width=1\textwidth]{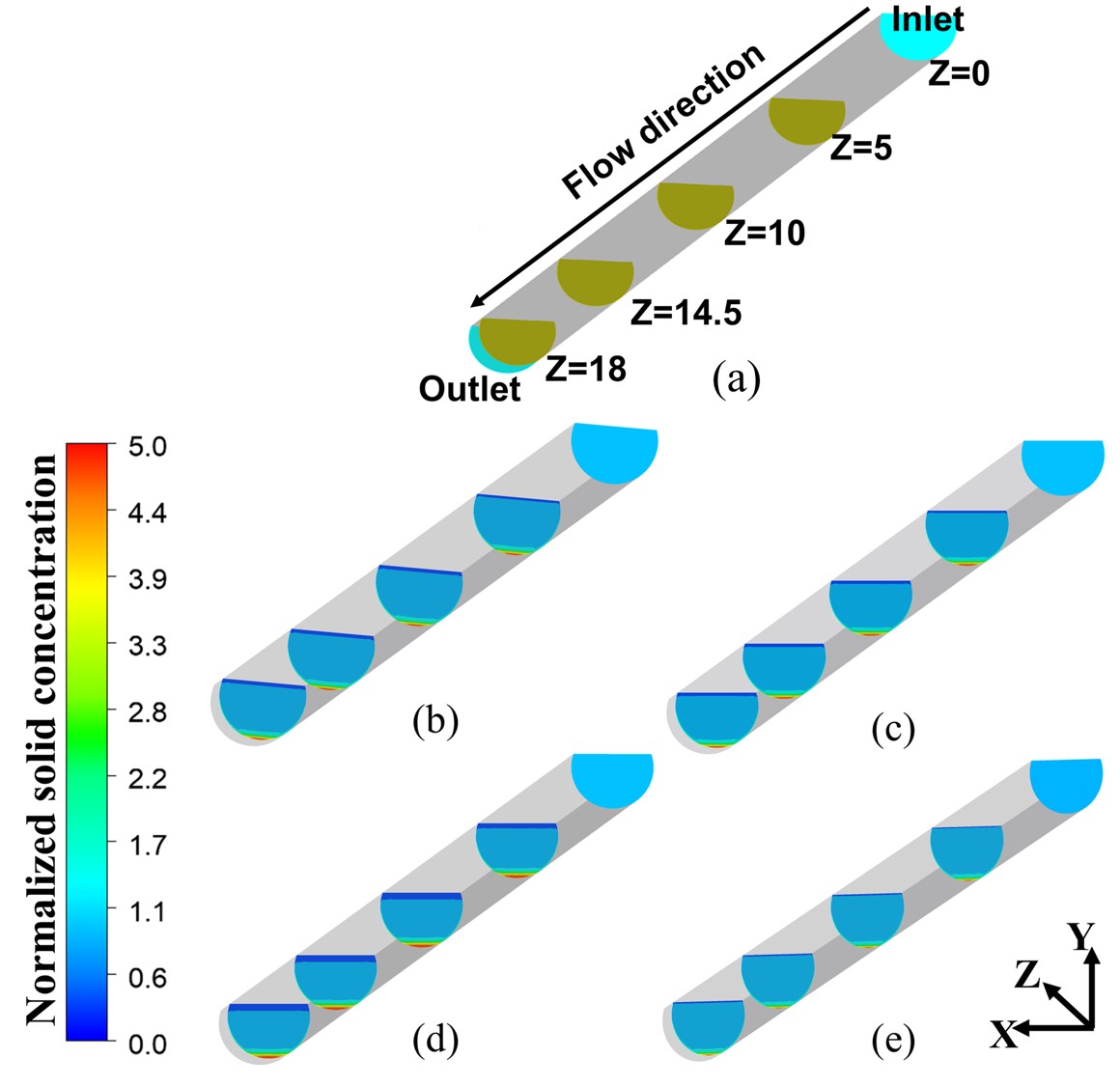}
\caption{(a) Planes of study along the length of the flume and contour representation of normalized solid distribution at distinct locations along the length of the flume for (b) PSD 1 (c) PSD 2 (d) PSD 3 (e) PSD 4 at bulk velocity 0.41 m/s and flume inclination 4\textdegree.}
\label{fig:Model8}
\end{figure}

The investigation of the flow behavior of multisize thickened slurry in a flume is performed using the CFD simulation. The parametric study is also presented to show the effect of the PSD, flume inclination, bubble size and bubble volume fraction on the settling behavior of thickened slurry in the flume. The analysis is performed at five different planes along the flume length as shown in Figure \ref{fig:Model8}(a). The planes are sectioned at the lengths Z = 0, 5, 10, 14.5, 18 m of the flume.

\subsection{Solids distribution}

\begin{figure}
	\centering
\includegraphics[width=0.95\textwidth]{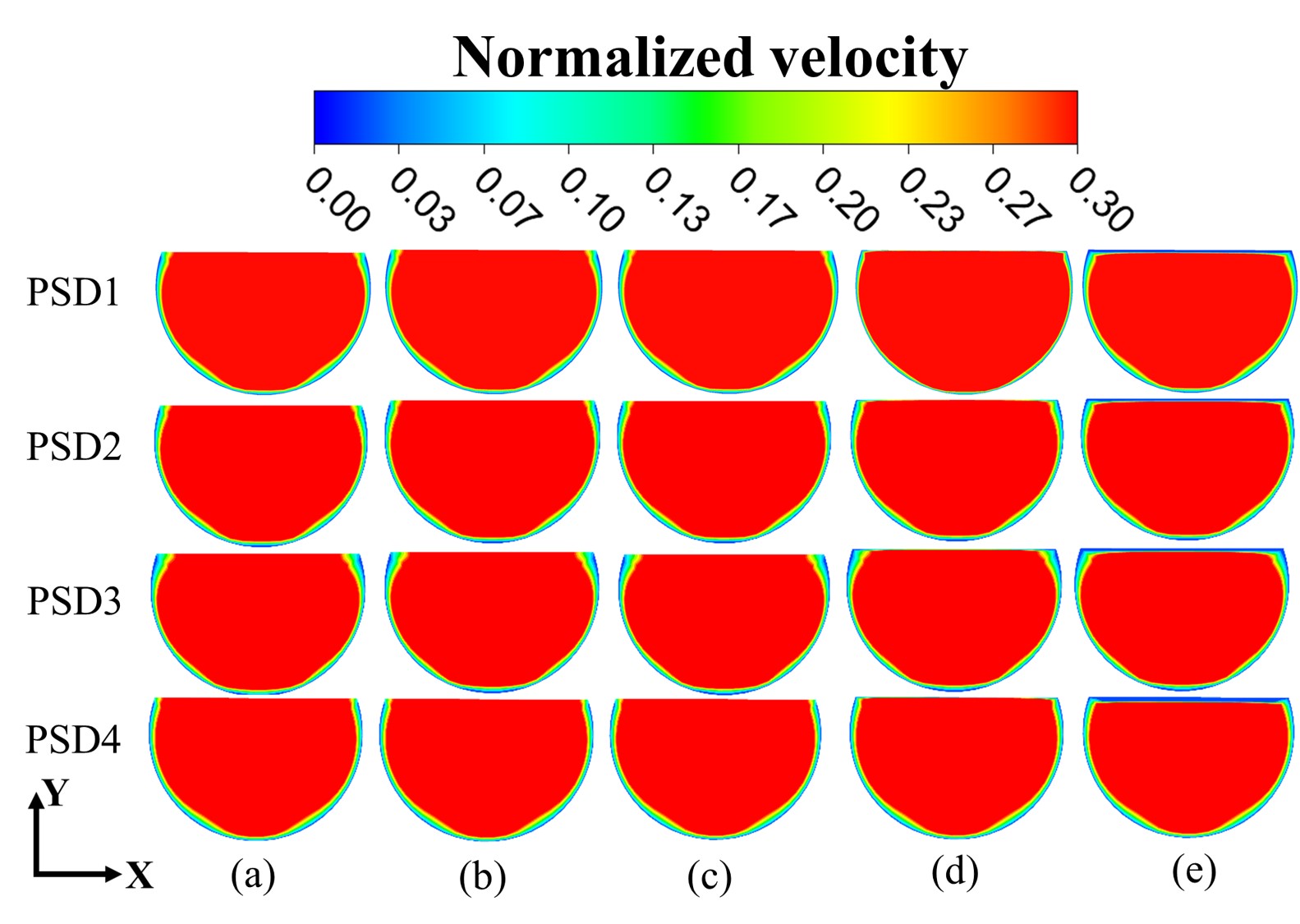}
\caption{Contour representation of normalized solid particle velocity at 14.5 m from the inlet of flume at different PSD for all range of solid particles (a) 75 µm (b) 105 µm (c) 149 µm (d) 210 µm (e) 296 µm at bulk velocity 0.41 m/s and flume inclination  4\textdegree.}
\label{fig:Model9}
\end{figure}

\begin{figure}
	\centering
\includegraphics[width=\textwidth]{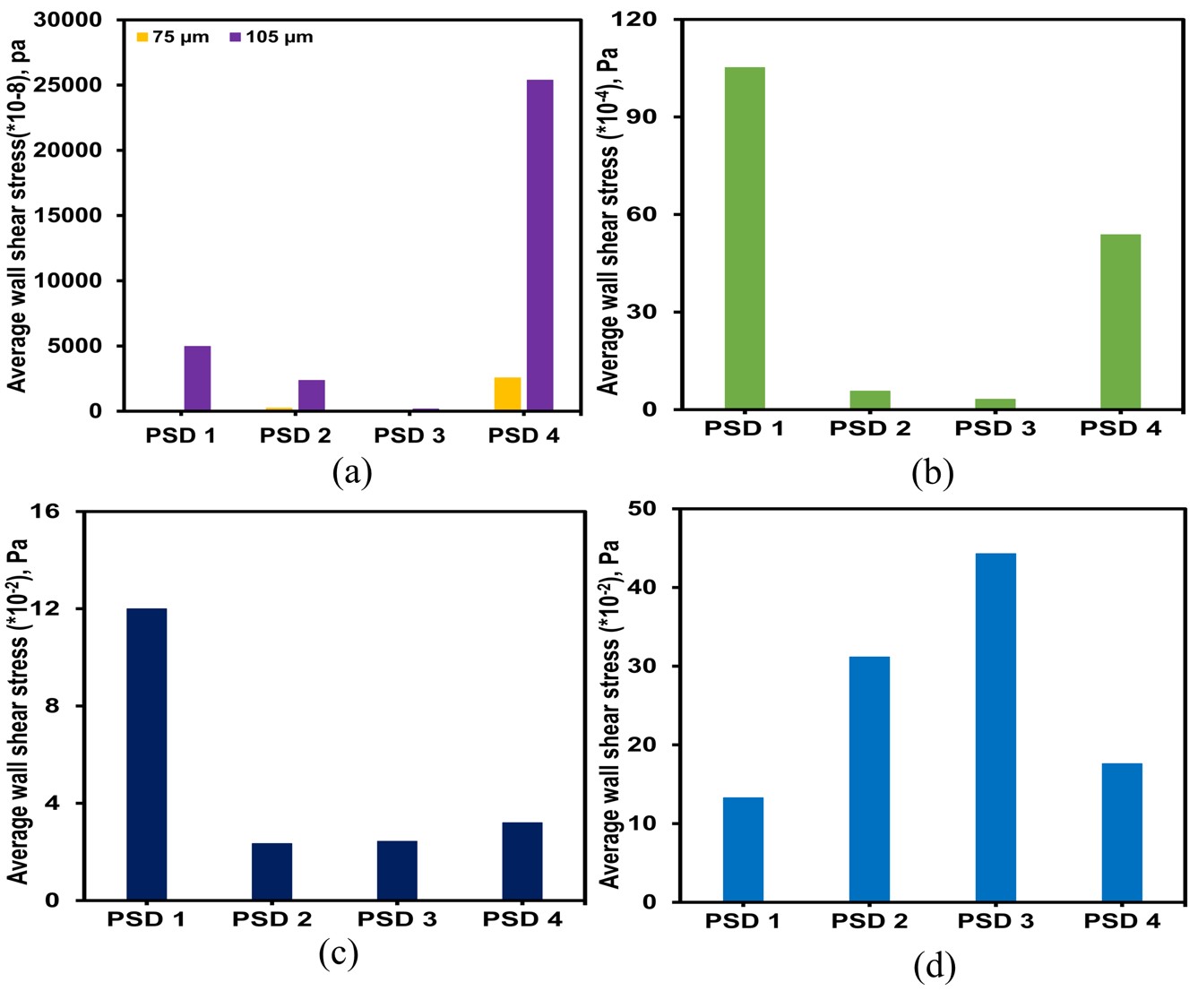}
\caption{\label{fig:Model11}~Variation of average wall shear stress over the flume wall with different PSDs at particle sizes 
(a) 75µm and 105 µm (b) 149 µm (c) 210 µm (d) 296 µm at bulk velocity 0.41 m/s and flume inclination 
  4\textdegree.}
\end{figure}

Figures \ref{fig:Model8}(b-e), \ref{fig:Model9}(a-e) and \ref{fig:Model11}(a-d) presented the effect of PSD inside the inclined flume during the thickened slurry transportation at the bulk solid volume fraction of 12.5\%. Figures \ref{fig:Model8}(b-e) depict the contour distribution of sand particles inside the flume for the different cases of PSD. The distribution of solids over the different cross-sectional plane of the flume along the length are represented in the normalised solid concentration terms, which is defined as the ratio of local solid volume fraction to the bulk solid volume fraction. The continuous settling of particles is found to increase towards the bottom section of the flume from inlet to outlet for all cases of PSD.
It is found from Figures $\ref{fig:Model8}$ (b-e) that for all cases of PSD, concentration changes in the range of about 0.01 to 5 in fraction from top to the bottom zone of the flume. Moreover, in the case of PSD 3, the difference is high and a settled particle dominating zone is found to be higher than in other cases.
\citet{thomas2004stabilised} reported that the particle settling at the invert section of the flume may have occurred due to a higher share rate over the identified zones.

The settling of particles results in the increase of the maximum settled concentration of sand and reduction of carrier fluid concentration at the bottom. Thus, the separation of fluid from the slurry may be improved because once the particle settles in the laminar flow regime, it is not possible to resuspend it \cite{paterson2004hydraulic}.            
Figures \ref{fig:Model9}(a-e) presented the velocity variation of solid particles over the plane at 14.5m from the inlet with different cases of PSDs for each particle size range at the bulk velocity 0.41 m/s and flume inclination 4°. The contour of velocity variation is shown in the form of normalized velocity as the ratio of local velocity to the bulk velocity. It can be seen from Figures \ref{fig:Model9}(a-e) that the velocity of particles increases from zero at the bottom to high at the top free stream surface. The velocity distribution contour confirms the model validation argument as well as the experimental observation of \citet{spelay2007solids}. Figures \ref{fig:Model9}(a-e) illustrate that the velocity distribution is same for all ranges of particles with different sets of PSD. However, some differences at the upper zone of the flume have also been seen in moving from particle size 149 µm to 296 µm. This may be attributed to the higher settling of coarser particles at the bottom, which results in the deficiency of particles at the upper zone of the flume.

Figures \ref{fig:Model11}(a-d) depict the average wall shear stress about the wetted perimeter of the flume for different cases of PSD at different range of particles with 0.41 m/s bulk velocity and flume inclination 4\textdegree. It can be found from the figures that the wall shear stress due to coarser particles is significantly higher than the finer particles for the all range of solid content variation. This significant resisting force of coarser particles along the flume wall may be observed due to its settling and sliding motion during transportation \cite{sharma2023assessment,sharma2022experimental}. Further, it can also be speculated from the wall shear stress distribution that the particle suspension, as well as the low interaction of the finer particle with the wall significantly reduced wall shear stress magnitude. Thus, it can be established from Figure \ref{fig:Model11}(c) that the high solid volume fraction with high segregation of solid 210 µm is responsible for the higher average wall shear stress with PSD1 compared to PSD3 even though both cases have the same range of 210 µm solid content in the slurry.  

\begin{figure}
	\centering
\includegraphics[width=1\textwidth]{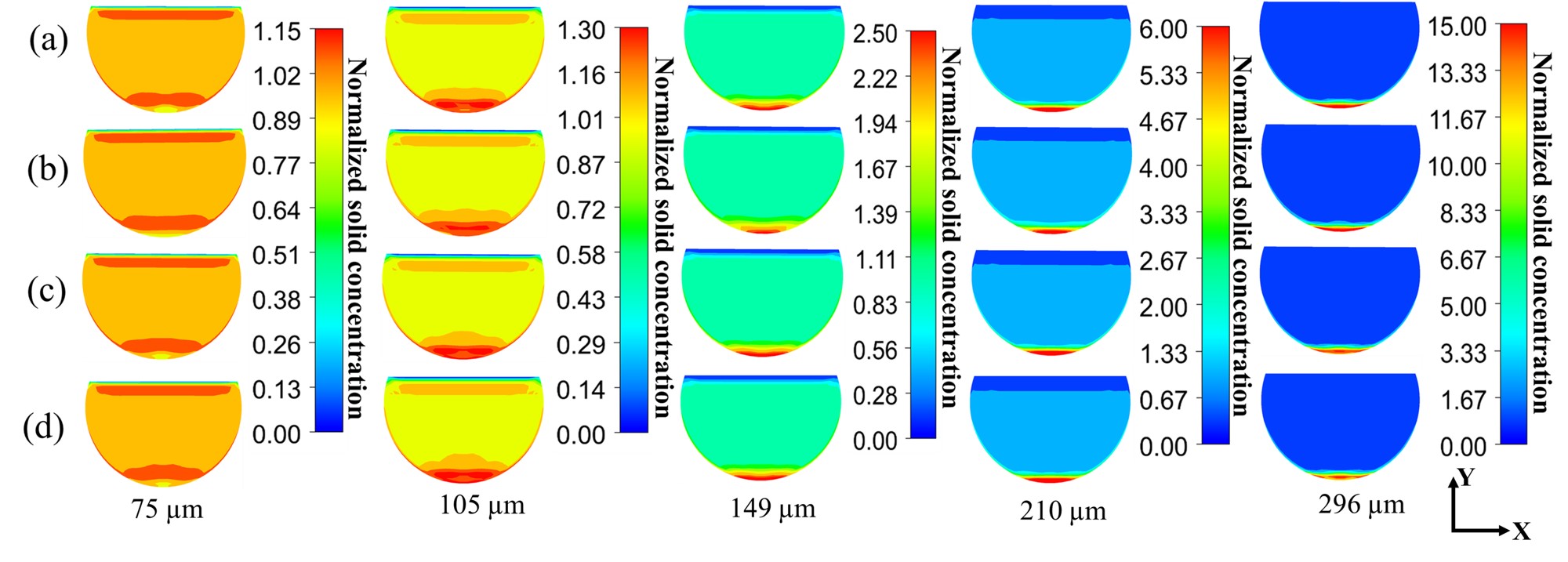}
\caption{\label{fig:Model12}~ Solid distribution over a cross-section plane 14.5 m from inlet for different particle sizes and flume inclination (a) 3° (b) 4° (c) 5° (d) 6° at PSD 1 and bulk velocity 0.41 m/s.}
\end{figure}
\subsection{Flume inclination}
In flumes slurry transportation is generally driven by gravity (inclination of channel), whereas in pipeline systems it is done by pressure gradient. Thus, the change of flume inclination may significantly change the dynamics of slurry flow. Figures \ref{fig:Model12}(a-d) depict the contour representation of normalized solid concentration for the four different inclination angles ranges 3°-6° over a plane at 14.5 m from inlet with PSD1 and bulk velocity as 0.41 m/s. It can be seen from Figures \ref{fig:Model12}(a-d) that at the constant depth of flow the settling behavior of the fine particles and coarse particles is different with the increase in inclination angle. With the increase in flume inclination angle the fine particles (75-149 µm) content is found to be increased in the sheared zone of slurry inside flume. This may have occurred due to a change in the direction of action of gravity in the flume bottom wall. The observed phenomenon emphasizes the settling of fine particles with the increase in flume inclination angle at the constant depth of the flow and volume flow rate of slurry. This observation is in-line with the work of \citet{sadeghi2023computational} in the case of poly-dispersed slurry flow in the inclined pipe. However, with the increase in particle size (210-296 µm) the segregation of particles at the flume invert section gets reduced with the increase in flume inclination angle, which may be due to the influence of carrier fluid velocity \cite{spelay2007solids}.

\begin{figure}
	\centering
\includegraphics[width=1\textwidth]{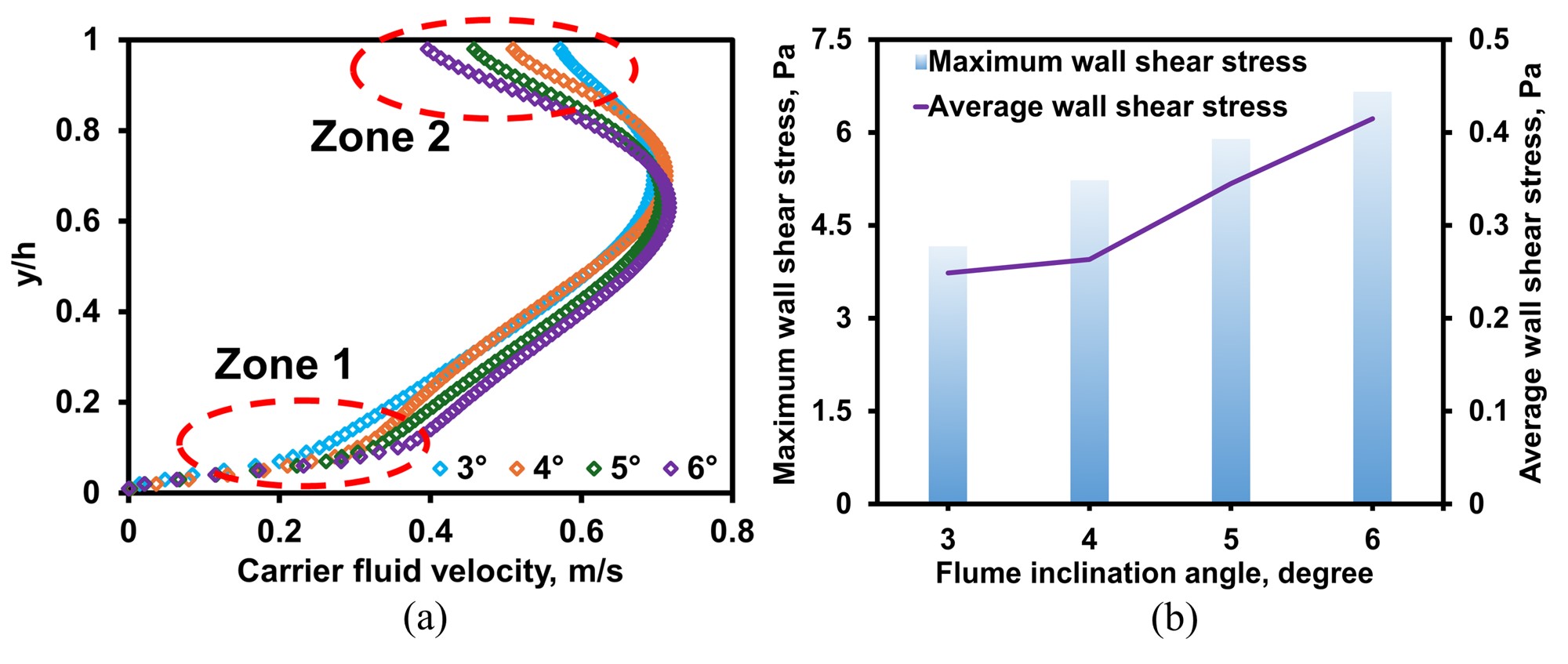}
\caption{\label{fig:Model13}~ (a) Carrier fluid velocity distribution over the plane 14.5 m from the inlet for different flume inclination angles with PSD 1, 12.5\% solid volume fraction and constant depth of flow (b) wall shear stress distribution over the flume wall for different flume inclination angles with PSD 1, bulk velocity 0.41 m/s and constant depth of the flow.}
\end{figure}

\begin{figure}
	\centering
\includegraphics[width=1\textwidth]{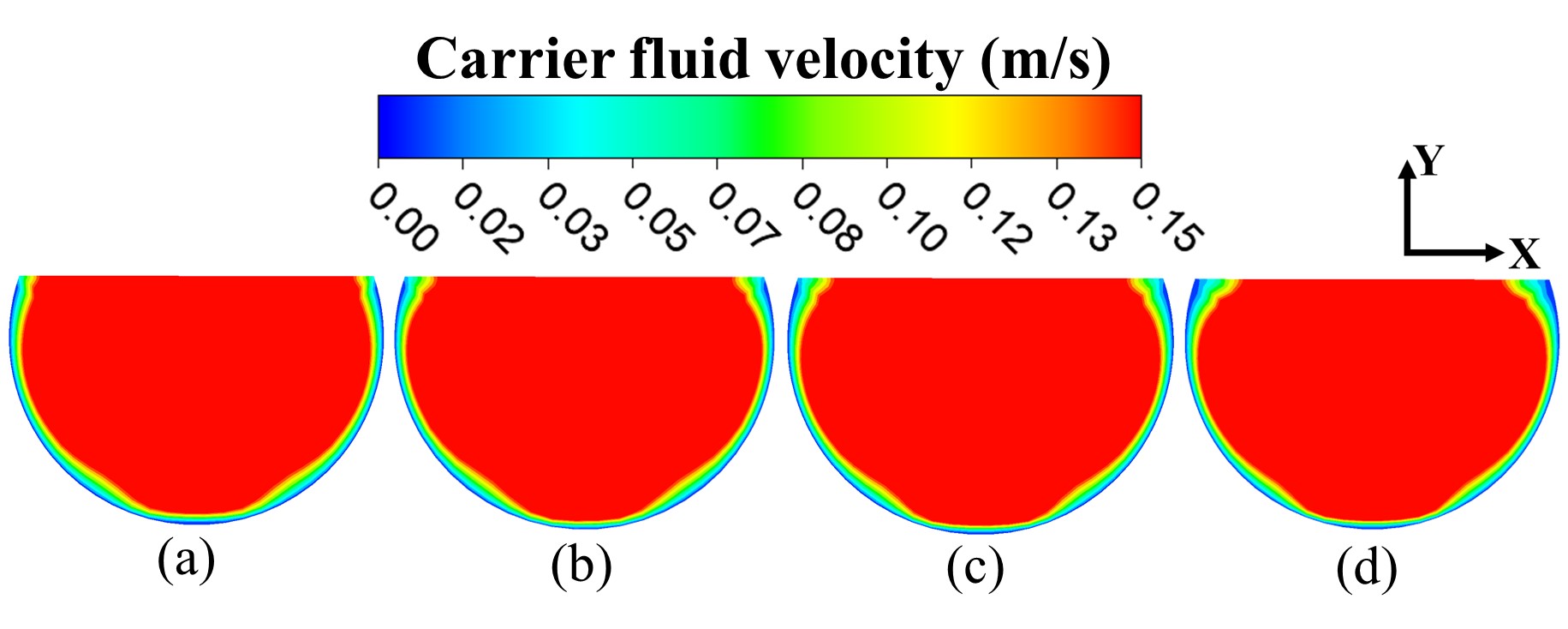}
\caption{\label{fig:Model14}~ Contour representation of carrier fluid velocity over the plane 14.5 m from the inlet for different flume inclination angles with PSD 1, 12.5\% solid volume fraction and constant depth of flow.}
\end{figure}
Figure \ref{fig:Model13}(a) shows the centre line carrier fluid velocity distribution over the same plane 14.5 m from the inlet for the different cases of flume inclination. The velocity distribution shows a significant difference in the velocity trend over the regions between 0.05~$<y/h<$~0.2 and 0.8~$<y/h<$~1 for all cases of flume inclination angle. It can be found from Figure \ref{fig:Model13}(a) that at zone 1 near to invert section of the flume, carrier fluid velocity is increasing with the increase in flume inclination while it shows decrement near the top section of the flume. The observed distribution confirms the argument of velocity influence on the segregation pattern of coarse particles at the flume invert section and fine particles at the top section of the flume. 
Figure \ref{fig:Model13}(b) presents the wall shear stress distribution with the change of flume inclination angle. It can be seen that the average and maximum wall shear stress over the wall of the flume is following the increasing trend with the flume inclination angle. The average wall shear stress at 6\textdegree~is found to be increased by about 1.6 times from the 3\textdegree~flume inclination. The variation of wall shear stress with flume inclination speculated the dominancy of particle settling effect over the fluid velocity for the thickened slurry flow with a constant depth of the flow. Moreover, the magnitude of average and maximum wall shear stress is found to be lesser than the yield stress of bingham fluid (thickened slurry), which also emphasizes the segregation of solid particles over the flume wall. Futhermore, the contour representation of carrier velocity is presented in the Figure \ref{fig:Model14}(a-d) to identify the fluid velocity distribution along the cross of the flume. It can be observed that qualitatively velocity distribution is not changing significantly along the cross-section.           
\subsection{Bubble size}
\begin{figure}[!ht]
	\centering
\includegraphics[width=\textwidth]{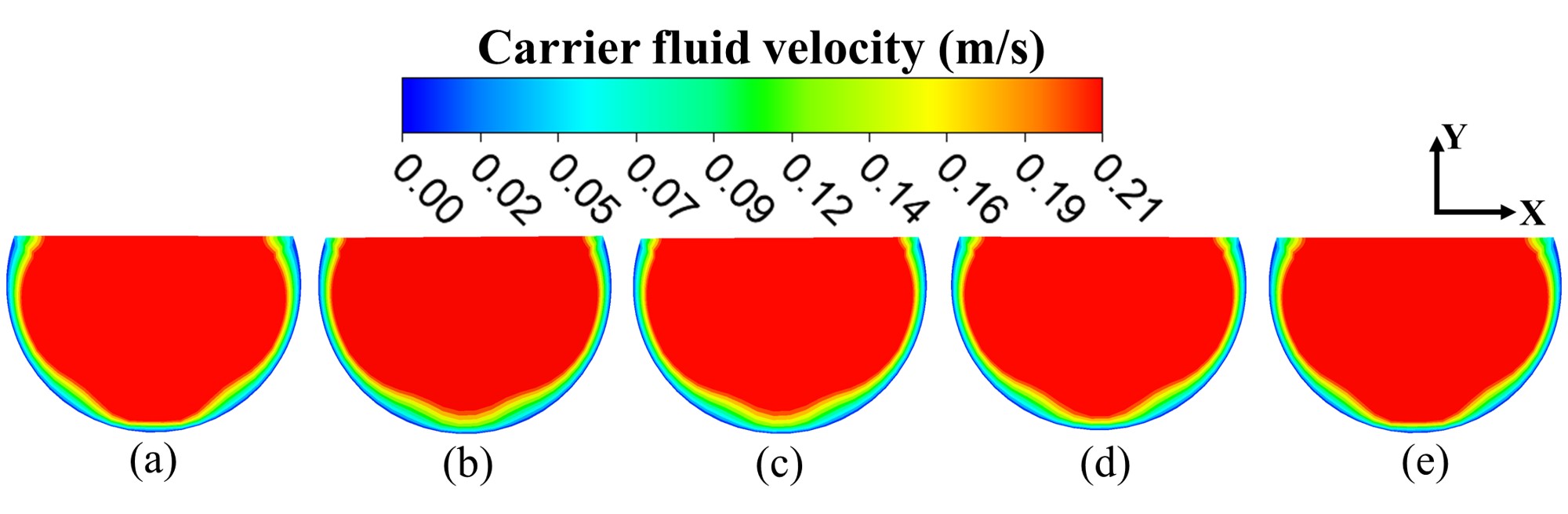}
\caption{\label{fig:Model15} Contour representation of carrier fluid velocity for different cases of bubble size (a) no bubble (b) 5 µm (c) 50 µm (d) 500 µm (e) 1000 µm at 0.0025 bubble volume fraction, bulk velocity 0.41 m/s, flume inclination 4° and PSD 1 over the plane of 14.5 m from the inlet.}
\end{figure}

\begin{figure}[!ht]
	\centering
\includegraphics[width=0.95\textwidth]{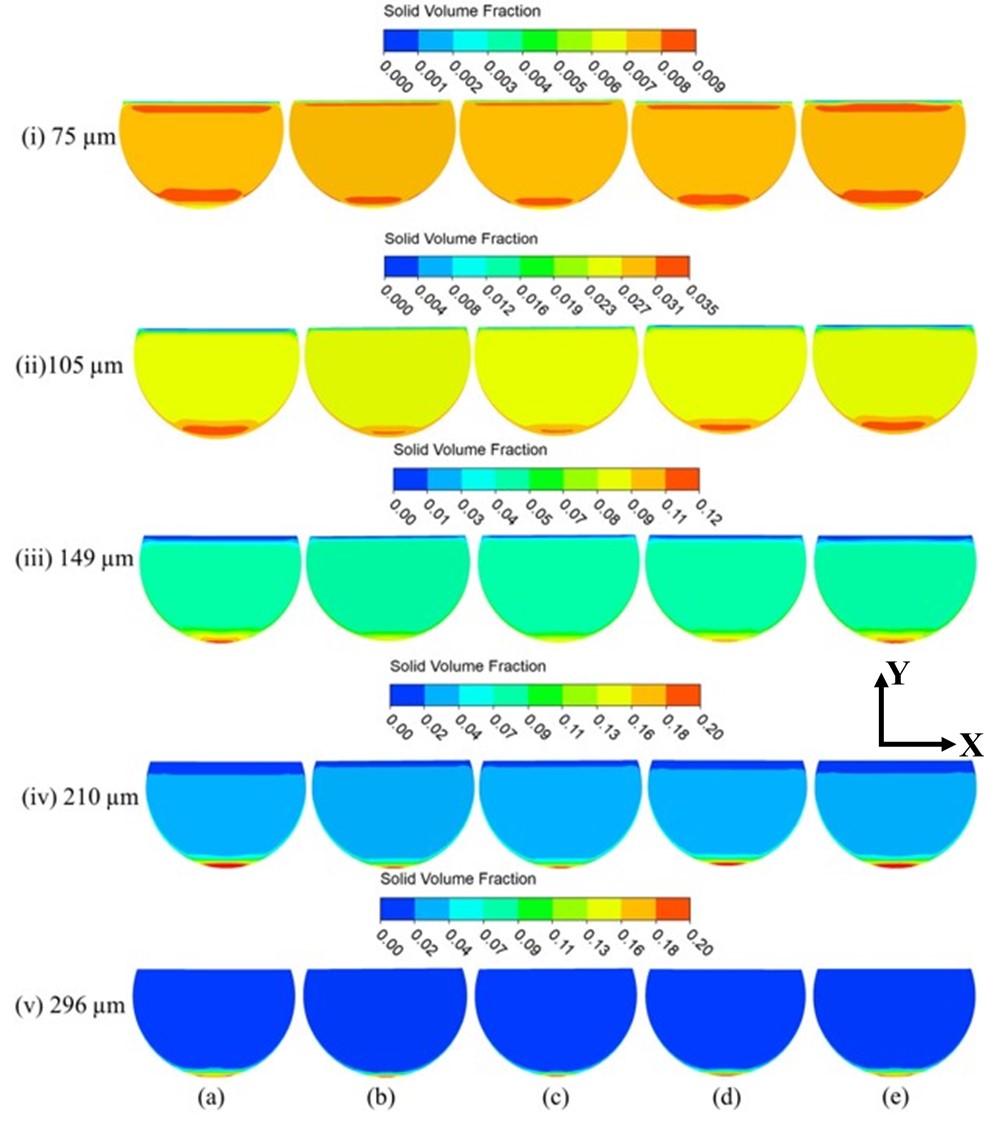}
\caption{\label{fig:Model16}~Contour representation of solid particle distribution for different particle sizes (i) 75 µm (ii) 105 µm (iii) 149 µm (iv) 210 µm (v) 296 µm with the variation of bubble size (a) no bubble (b) 5 µm (c) 50 µm (d) 500 µm (e) 1000 µm at 0.0025 bubble volume fraction, bulk velocity 0.41 m/s, flume inclination 4\textdegree and PSD 1.}
\end{figure}

This section reported the effect of gas bubble size on the settling behavior of the multi-size particulate slurry inside the inclined flume. The bubble size ranging from 5 µm to 1000 µm \cite{sontti2023computational} for the fixed condition of the slurry system at 0.0025 bubble volume fraction, bulk velocity 0.41 m/s, flume inclination 4° and PSD 1 with 12.5\% solid volume fraction. Figures \ref{fig:Model15} and \ref{fig:Model16} depict the contour representation of carrier fluid velocity and solid distribution over the plane 14.5 m from the inlet of the flume.\

Figures \ref{fig:Model15}(a-e) show the insignificant change in carrier fluid velocity with the variation of bubble size. However, the presence of bubbles inside the flume from finer to coarser range influences the solid distribution. It can be seen from Figures \ref{fig:Model16}(i-v)(a-e) that the bubble occupancy affects the multi-size solids settling. The contours clearly show that with finer size bubbles (5 and 50 µm), solids of all ranges are significantly suspended and will alter the settling phenomenon over the invert section of the flume. The observed change in physics of particle settling may be observed due to the lack of recoalescence of the finer gas bubbles due to presence of solid particles \cite{rosas2018measurements}. Further on increasing the bubble size from 500 µm to 1000 µm, the evolution of the solids tends to increase at the bottom of the flume. The particle settling trend for all range of solids at 1000 µm is found to be very similar with no bubble occupancy case. This may be due to the dominancy of the buoyant force in the case of large bubbles, which results in the separation of bubbles towards the top free surface and particles towards the invert side of the flume. The predicted numerical findings are also agreed with the work of \citet{rosas2018measurements} and \citet{sontti2023computational}. The predicted results signify that the occupancy of finer-sized bubble in the slurry results in the dissipation of solids inside the flume.\

Furthermore, the chord average solid volume fraction is estimated over the same plane to quantify the zone of particle dissipation due to the presence of bubbles. Figure S1 illustrates the variation of chord average solid volume fraction from the bottom of the flume to the top free surface. Due to the bubble entrapment inside the flume system, the particle settling got shattered more importantly near the bottom section of the flume. However, on increasing the bubble size the solids settling seems to be improved. The zone inside the flume in between 0 $<y/D<$ 0.1 shows a significant variation in the particle solid volume fraction due to bubble entrapment. This might have been due to decreased bubble attachment efficiency with the increase in the settling velocity of particles \cite{ma2023effect, jing2021settling}. The continuous increment in settling velocity results in the free and static downward movement of particles, thus, the bed level is found to be increased toward the invert section for the case of bigger size bubbles. The finding implies that the presence of bigger size bubbles inside the open channel slurry transportation system may enhance the settling phenomenon of multi-size solids which could be a useful understanding for the dewatering process of thickened slurry.

\subsection{Bubble volume fraction}
\begin{figure}[!ht]
	\centering
\includegraphics[width=\textwidth]{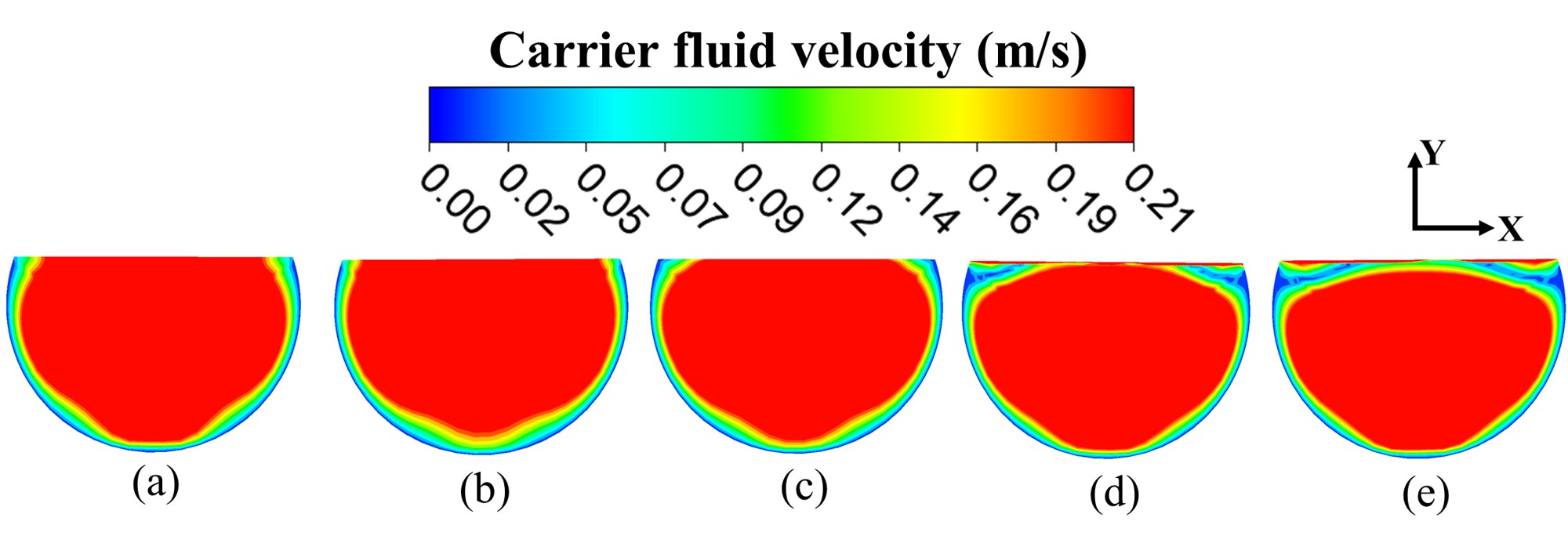}
\caption{\label{fig:Model17}~Contour representation of carrier fluid velocity at the plane 14.5 m from the inlet for bubble volume fractions (a) no bubble (b) 0.0025 (c) 0.01 (d) 0.02 (e) 0.03 at 50 µm bubble size, bulk velocity 0.41 m/s, flume inclination 4° and PSD 1. }
\end{figure}

\begin{figure}[!ht]
	\centering
\includegraphics[width=0.9\textwidth]{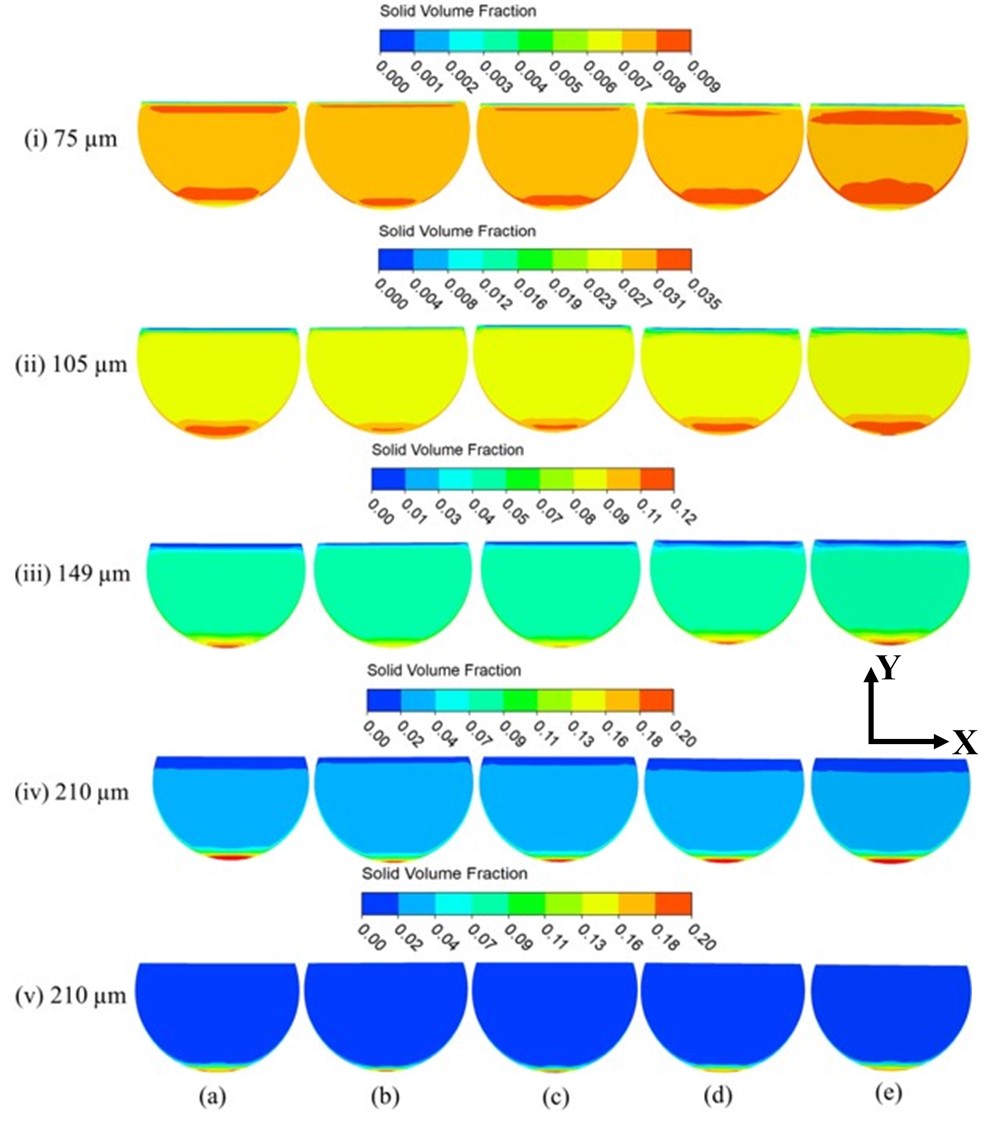}
\caption{\label{fig:Model18}Solid particle distribution over a plane at 14.5 m from inlet of the flume for different particle sizes (i) 75 µm (ii) 105 µm (iii) 149 µm (iv) 210 µm (v) 296 µm with the variation of bubble volume fraction (a) No bubble (b) 0.0025 (c) 0.01 (d) 0.02 (e) 0.03 at 50 µm bubble size, bulk velocity 0.41 m/s, flume inclination 4° and PSD 1.}
\end{figure}

The effect of bubble entrapment in the slurry flow with different range of bubble concentrations is reported in this section. Figures \ref{fig:Model17}(a-e) and \ref{fig:Model18}(i-v)(a-e) illustrate the variation of carrier fluid velocity and solid distribution, respectively, with the variation of bubble volume fraction ranges 0.0025 to 0.03 at 0.41 m/s bulk velocity, PSD 1 and 4° flume inclination over the plane 14.5 m from the inlet. \


Figures \ref{fig:Model17}(a-e) show that the carrier fluid shifts towards the bottom of the flume due to influence of bubble volume fraction, which results in the lowering of the magnitude of velocity near the free surface. The identified flow behavior may be attributed to the accumulation of bubbles near the top free surface of the flume. This numerical predicted slurry flow behavior with the inception of significant bubble volume fraction is very similar with the observation of \citet{sontti2023computational}. Moreover, Figures \ref{fig:Model18}(i)(a-e) depict the increment in the accumulation of fine particles near the top free surface of the flume with bubble volume fraction, which further enhances the zone of relatively low velocity near the top free surface. This might have occurred due to increased attachment efficiency of fine particles and bubbles at higher bubble volume fraction, which leads to the upward movement of particles \cite{ma2023effect}.

Furthermore, to identify the pattern of solid distribution with the presence of bubbles, the chord average solid volume fraction against the depth of the flow at the same operating condition is presented in Figure S2. The total solid content variation trend agrees with the individual solid settling phenomenon as illustrated in Figures \ref{fig:Model18}(i-v)(a-e). Moreover, the solid distribution with the variation of bubble volume fraction is also very similar with the distribution trend at different bubble sizes. The increase in the bubble volume fraction leads to the decrease in mixture density of the carrier fluid and bubble and an increase in terminal settling velocity of solid particles \cite{jing2021settling}. Thus, the dominancy of settling velocity increment over the lift force created by bubbles causes more settling of solids at higher bubble volume fractions than at low bubble volume fractions.

\section{Conclusion}
In this work, 3D unsteady numerical modelling of a flume with multisize particulate slurry using granular E-E modelling has been performed. The numerical model is validated by comparing the predicted and measured chord average solid volume fraction over the plane 14.5 m from the inlet. The CFD predicted results showed a reasonable agreement with experimentally measured data within the uncertainty range ±15\% and ±5\% for solid volume fraction and average carrier fluid velocity, respectively. A detailed sensitivity analysis for the drag model selection predicts that the symmetric model is a precise option for modelling carrier-solid interaction.

The performed parametric study established the effect of PSD, flume inclination and bubble influence on the settling behavior of solid particle. PSD variation ranges 75-296 $\mu \mathrm{m}$ notably influences the settling behaviour of slurry solids inside the open channel system. Moreover, this settling increases wall shear stress in the range of 9.87e-08  to 0.44 Pa for 75 $\mu \mathrm{m}$ to 296 $\mu \mathrm{m}$ size particles, respectively. The fine particles settled more with the increase in flume inclination angle while the dissipation of coarse particles increases with the increase in flume inclination from 5° to 6°. Inception of the coarser bubble found to lead the segregation of solids, while finer bubbles cause the dissipation of solids inside the flume.

The presented numerical study provided an in-depth understanding of the settling behavior of solids during the transportation of non-newtonian thickened slurry through a flume. This work may guide the optimization of slurry deposition to maximize the fine capture. Furthermore, the outcomes of the study can be used for the design and development of an effective dewatering system of thickened slurry in the open channels, thus, the environmental concern by slurry transportation might be resolved.

\section*{Supplementary Material}
\noindent A supporting document is available on variation of chord average solid content at different bubble size over a cross-section  and variation of chord average solid content at different bubble volume fraction over a cross-section.

\section*{Credit author statement}

\noindent \textbf{S. Sharma:} Conceptualization (lead); Methodology (lead); Planned and performed the simulations (lead); Model validation (lead); Formal analysis (lead); software (lead); Visualizations (lead); Writing--original draft (lead); Writing--review \& editing (lead). \textbf{S. G. Sontti:} Visualizations (supporting), Writing--review \& editing (supporting).  \textbf{W. Zhang:} Writing--review \& editing (supporting). \textbf {P. Nikrityuk:} Writing--review \& editing (supporting) , \textbf{X. Zhang:} Conceptualization (supporting); Methodology (supporting); Data interpretation (lead); Project administration(lead); Writing--review \& editing (lead); Resources (lead); Supervision (lead).

\section*{Declaration of competing interest}
\noindent The authors declare that they have no known competing financial interests or personal relationships that could have appeared to influence the work reported in this paper.

\section*{Acknowledgement}
\noindent 
The authors are thankful to the support from Canada Research Chairs Program, Discovery Project, Alliance Grant from Natural Sciences and Engineering Research Council of Canada. We also thank the Digital Research Alliance of Canada (https://alliancecan.ca/en) for continued support through regular access of high-performing server HPC cedar and Graham Cluster. 
 This work was supported by Imperial Oil Limited and Alberta Innovates through the Institute for Oil Sands Innovation at the University of Alberta (IOSI).


\section*{Data availability}
\noindent The data supporting this study's findings are available from the corresponding author upon reasonable request.

\clearpage
\section*{Nomenclature}
\subsection*{Latin Symbols}
\begin{tabbing}
    $D$ \hspace{2cm} \= Pipe diameter (L) \\
    $d$ \> Particle diameter (L) \\
    $p_i$ \> Predicted solid volume fraction \\
    $m_i$ \> Measured solid volume fraction \\
    $d_{50}$ \> Median particle diameter \\
    $y$ \> Depth of flow \\
    $h$ \> Maximum depth of flow \\
    $g$ \> Gravitational acceleration (L T$^{-2}$) \\
    $g_0$ \> Radial distribution function (--) \\
    $P$ \> Locally-averaged pressure (M L$^{-1}$ T$^{-2}$) \\
    $t$ \> Time (T) \\
    $v$ \> Velocity of phases (L T$^{-1}$) \\
    $F_{\mathrm{drag}}$ \> Drag function (--) \\
    $\beta$ \> Aggregation kernel function \\
    $x$ \> Horizontal coordinate (L) \\
    $z$ \> Axial coordinate (L) \\
    $e$ \> Restitution coefficient (--) \\
    $I$ \> Second invariant of the deviatoric stress tensor (--) \\
    $\norm{\vec{v}_{si}^{\,\prime}}$ \> Fluctuating solids velocity (L T$^{-1}$) \\
    $K_{fi}$ \> Momentum exchange coefficient between fluid and solid phase (--) \\
\end{tabbing}

\subsection*{Greek Symbols}
\begin{tabbing}
    $\alpha$ \hspace{2cm} \= Locally-averaged volume fraction (--) \\
    $\mu$ \> Dynamic viscosity (M L$^{-1}$ T$^{-1}$) \\
    $\rho$ \> Density of phases (M L$^{-3}$) \\
    $\phi_{fsi}$ \> Energy exchange between fluid and solid phases (E) \\
    $\tau$ \> Shear stress (M L$^{-1}$ T$^{-2}$) \\
    $\dot{\gamma}$ \> Shear strain rate (T$^{-1}$) \\
    $\alpha_{s,\mathrm{max}}$ \> Maximum packing limit (--) \\
    $\Theta$ \> Granular temperature (L$^{2}$ T$^{-2}$) \\
    $\varphi$ \> Angle of internal friction (--) \\
\end{tabbing}

\subsection*{Subscripts}
\begin{tabbing}
    $f$ \hspace{2cm} \= Fluid \\
    $s$ \> Solid \\
    $si$ \> Solid particles \\
    $p$ \> $p^{\text{th}}$ phase \\
\end{tabbing}

\bibliography{mybibfile}
\end{document}